\documentclass[prc,preprint,superscriptaddress,showpacs,nofootinbib,floatfix]{revtex4}
\usepackage{epsfig,bm,feynmf}
\usepackage{graphics}
\usepackage{graphicx}
\usepackage{subfigure}
\usepackage{amsmath,amssymb}
\usepackage{mathrsfs}

\def\dis{distribution}
\def\pt{p_T}
\def\ppt{$p_T$}
\def\bq{\begin{eqnarray}}
\def\eq{\end{eqnarray}}
\begin{document}

\title{A unified study of the production of all identified hadrons over wide ranges of transverse momenta at LHC}
\author{Lilin Zhu}
\email{zhulilin@scu.edu.cn}
\affiliation{Department of Physics, Sichuan University, Chengdu 610064, China}

\author{Rudolph C. Hwa}
\email{hwa@uoregon.edu}
\affiliation{Institute of Theoretical Science and Department of Physics, University of Oregon, Eugene, OR 97403-5203, USA}

\begin{abstract}
The production of all identified hadrons at the CERN Large Hadron Collider (LHC) is studied with emphasis on the $p_T$ distributions up to 20 GeV/c in central collisions at $\sqrt{s_{NN}}=2.76$ TeV. The parton recombination model is used to determine the hadronic \ppt\ spectra from the quark \dis s. From the heavy hyperon spectra it is known from earlier studies that the $u, d, s$ thermal \dis s in \ppt\ are exponential with large inverse slopes that cannot be identified with any temperature in conventional fluid models. They are used as inputs in our model together with shower partons determined from our treatment of momentum degradation that uses high-\ppt\ pion data as input. Those thermal and shower partons are used to calculate the $p_T$ \dis s of all observed hadrons ($\pi, K, p, \Lambda$, $\Xi$, $\Omega$ and $\phi$) over wide ranges of \ppt, so the system is highly constrained. We show how well the LHC data can be reproduced 
with only a few parameters to adjust. Centrality dependence has not been studied. What is learned is that minijets are important, not only in giving rise to abundant shower partons, but also in the conversion of semihard partons in the medium to soft partons that enhance the thermal partons. Since the conversion process can occur throughout the expansion phase of the high-density medium, this work provides the basis for questioning the validity of the assumption of rapid equilibration.
\end{abstract}
\maketitle

\section{Introduction}

In a recent paper  a remarkable property is reported about the transverse momentum (\ppt) spectra of baryons produced in heavy-ion collisions over wide ranges of collisions energy, \ppt\  and centrality \cite{Hwa:2018qss}. It was found that, when an appropriate function of the baryon distribution is plotted against \ppt, the data exhibit nearly perfect exponential behavior regardless of strangeness content; moreover, the corresponding inverse slopes ($T_h$) can be simply related across energy $\sqrt{s_{NN}}$, strangeness number ($n_s$) and the number of participants $N_{\rm part}$. Furthermore, the values of $T_h$ are much larger than the temperatures usually discussed in the conventional hydrodynamical description of the plasma medium \cite{hydro1, hydro2} --- by a factor of 3 or so. Those are phenomenological observations that are independent of any model and are urgently in need of explanation.

The fact that the inverse slopes \ppt\ are much larger than any temperature that has been discussed in any previous models and the associated exponential behavior being valid over wider ranges of \ppt\ than has ever been realized before suggests the existence of some dynamical process at work that has alluded recognition. 
Our suggestion in \cite{Hwa:2018qss}  is that there are minijets produced in the collisions, which are dissipated in the medium and enrich the soft partons in ways that contradict the assumption of rapid equilibration. In this paper we follow up on that notion and expand our investigation to include the \ppt\ spectra of all observed identified hadrons, including pions and strange mesons, i.e., $\pi, K, \phi, p, \Lambda$, $\Xi$ and $\Omega$, in a unified study of all partonic processes that can contribute to  the formation of those hadrons in the recombination model \cite{hy1,hy2,hz1,hz2}.

Generally speaking, theoretical study of hadron production in heavy-ion
collisions at high energies is usually separated into different
camps, characterized by the regions of transverse momenta $p_T$ of
the produced hadrons. At low $p_T$ statistical hadronization and
hydrodynamical models are generally used \cite{pbm, ph, kh, tat, gjs}, whereas at high
$p_T$ jet production and parton fragmentation with suitable
consideration of medium effects in perturbative QCD are the
central themes \cite{pq1, pq2, ia, mt, jet}. In the past the two approaches have been studied
essentially independent of each other with credible success in
interpreting the data, since their dynamics are decoupled at the
energies investigated. The situation has changed at the CERN
Large Hadron Collider (LHC), where 
thousands of soft hadrons are produced on the one hand, and multiple hard jets on the other.
Minijets that are copiously produced at intermediate $p_T$ can
fragment into soft partons with multiplicities so high that their
effects on the hadronization of all partons created in the soft sector cannot be ignored. 
In order to find a unified explanation of the production of hadrons of all species for all $p_T$ measured up to 20 GeV/c, we shall focus on Pb-Pb collisions at $\sqrt{s_{NN}}=2.76$ TeV for which data are available online \cite{Adam:2015kca, kslambda, ba2}.

Hard parton scattering and hydrodynamical flow are processes that
involve very different time scales. It would be hard to
incorporate them into a unified formalism that describes all
aspects of the system, including thermalization time, initial configuration, 
fluid nature of the medium, its quenching effect on the hard and semi-hard 
partons, the creation of shower partons, and the hadronization of
all partons at the end of the whole process. Our attempt here is
far from being so ambitious. We focus only on the $p_T$
dependencies of the hadrons produced in the interval 0.5 to 20 GeV in a
formalism that can be valid throughout that range, provided that
we use some model inputs for the thermal component of the low-$p_T$ behavior to supplement
the hard component that can be calculated at high $p_T$. We use quark recombination
to treat hadronization, applied uniformly at all $p_T$. In
treating the degradation of momenta of hard and semihard partons (which are not calculable reliably) 
we shall adjust some parameters to fit the high-$p_T$ data. Since
we aim to confront the $p_T$ spectra of all observed hadrons,
 the system is highly constrained. 

Our investigation of produced hadrons with various contents of
strangeness also reveals contrasting features of heavy-ion physics
not commonly addressed. Whereas hard scattering of gluons and
light quarks can readily occur at high energies, jet fragmentation
into multi-strange hadrons like $\Omega$ and $\phi$ is rare even
at LHC. But the production of $\Omega$ relative to $p$ grows
exponentially with $p_T$ even to the highest $p_T$ measured around 7 Gev/c. The data for that will be exhibited explicitly in Sec.\ II, along with several other pieces of data presented in novel ways so as to emphasize the problems that have not been commonly discussed. 
Surely, one cannot expect $\Omega$ to be easily produced at
$p_T=7$ GeV/c by jet fragmentation. 
In the framework of the recombination models \cite{hy1,hy2,hz1,hz2} the baryons are formed from quarks with $p_i$ in the neighborhood of 2-3 GeV/c, which can be populated by the fragmentation of minijets that are abundonly produced at LHC. Those quarks, light and strange, are generated in the medium and become a part of the thermal component before hadronization. 

We do not consider here other features of heavy-ion collisions
besides $p_T$ distributions, most notably the azimuthal dependence
in non-central collision. Conventional description of elliptic
flow does not consider the effects of jets. We shall treat that
subject separately, after our concern about the shower partons
establishes a footing in the general terrain of heavy-ion physics.

To clarify the nature of our approach it is necessary to contrast it from the standard model based on hydrodynamics. If hard and semihard partons produced in high-energy nuclear collisions are important in their effects on soft particles, then one should recognize that their in-medium radiated daughter partons take some time to thermalize, much longer than the rapid equilibration time ($\tau_0\sim 0.6$ fm/c) usually assumed in hydro calculations. A hard parton produced near the center of the medium in central collisions would take about 6 fm/c to reach the surface. Thus rapid thermalization is not realistic if minijets are important, as we shall show that they are at LHC. As a consequence, we cannot make use of hydro results in our approach, nor can hydro results be used to censure our calculations. For example, the thermal parton \dis\ that we consider is not to be identified with any \dis\ of the fluid constituents in the hydro medium. Also, in the hydro treatment $v_2$ is identified with elliptic flow, but it is only a possible, not a necessary, explanation. Other explanations are also possible; see, for example, Refs.\ \cite{h08,chy,hz3}. In this paper we consider only central collisions and establish the importance of shower partons recombining with the thermal partons. It is suggested that a reader withholds comparison with hydro treatment until the main points advanced here can be made.

This paper is organized as follows: In Sec.\ II we show experimental features that involve pions as well as baryons so as to find clues and to raise pertinent questions. Section III describes the general formulation of our approach to the problem. Shower parton distributions are discussed in detail in Sec.\ IV with emphasis on how the degradation of parton momenta is treated. With those partons shown to be important even in the intermediate transverse-momentum region, the recombination of shower partons from nearby jets becomes a possibility that is considered in Sec.\ V. With all the basic inputs on partons at hand we then proceed to the determination of the transverse-momentum distributions of $\pi, p, K$ and $\Lambda$ in Sec.\ VI. Multi-strange hyperons and meson are treated in Sec.\ VII with detail equations given in the Appendices. Section VIII contains our conclusion.

\section{Notable Experimental Features}
We consider some data from LHC that can be taken to suggest
something unusual about the usual observables. Compared to the
data at RHIC energies and below, it seems that simple
extrapolation to Pb-Pb collisions at 2.76 TeV is likely to miss
some new physics.

With the charged-particle multiplicity density (i.e., per participant pair) averaged over $|\eta|<0.5$ defined as
\bq
\rho_{\rm ch}^{AA} = dN_{\rm ch}^{AA}/d\eta |_{|\eta|<0.5}/(\langle N_{\rm part}\rangle/2) ,   \label{1}
\eq
it is found  for central Pb-Pb collisions  at $\sqrt{s_{NN}}=2.76$ TeV that 
\bq
\rho_{\rm ch}^{\rm Pb-Pb}(\sqrt{s_{NN}}=2.76)\ ^>_\sim\ 2 \rho_{\rm ch}^{\rm Au-Au}(\sqrt{s_{NN}}=0.2) ,   \label{2}
\eq
i.e.,  more than twice larger than the value measured in Au-Au collisions at $\sqrt{s_{NN}}=200$ GeV  \cite{ka, ATLAS:2011ag}. What is
notable is that in a semilog plot of $\rho_{\rm ch}^{\rm AA}(\sqrt{s_{NN}})$ versus ${\rm log}\sqrt{s_{NN}}$
 a straight line can be drawn through all the points from $\sqrt{s_{NN}}=2.5$ GeV to 200 GeV, but its extension to 2.76 TeV misses badly the LHC data point, which is more than 30\% higher than the value on the linear extrapolation in ${\rm log}\sqrt{s_{NN}}$. 
The dramatic increase above
the logarithmic dependence suggests the onset of new physics.

Another difference between LHC and RHIC is the dependence on $p_T$.
From the $p_T$ distributions measured at the two energies, 2.76
and 0.2 TeV, we can calculate their ratio. When the data points
are not in the same $p_T$ bin, we make Lagrangian interpolation between
adjacent bins in the RHIC data \cite{aa, xlz} to match the LHC bin \cite{Adam:2015kca, kslambda, ba2}. The
result for pion is shown by the solid (black) line in Fig.\ 1. Note the
exponential increase by two orders of magnitude as $p_T$ is increased 
up to 10 GeV/c. Similar increases are noted for
 $p$ and $\Omega$ up to $p_T\sim 6$ GeV/c.  It is not unexpected that
more high-$p_T$ particles are produced at higher collision energy and that pions numerically dominate all other species. But Fig.\ 1 shows that for each identified specie of particles there are abundantly more particles produced at LHC relative to RHIC at higher \ppt; in fact, the increases of the ratios up to $\pt \sim 6$ GeV/c are approximately the same independent of the species.
The question is what effects do the hard scatterings of partons
have on the production of intermediate-$p_T$ hadrons at $2<p_T<6$
GeV/c. Furthermore, it is reasonable to ask whether the physics at
low $p_T$ can be treated by hydrodynamics as at RHIC, totally
decoupled from the physics at high $p_T$. If jets are copiously
produced in order to account for a factor of $10^2$ at
$p_T\sim10$ GeV/c in Fig.\ 1, why would their fragmentation
products not populate the low-$p_T$ region below 2 GeV/c? Our
knowledge on fragmentation functions derived by leptonic
collisions tells us that the distribution of hadronic products
increase monotonically with decreasing momentum fraction \cite{kkp}.
What we learn from Fig.\ 1 is that the similarity of the behaviors of $\pi, p, \Omega$ in the region $0.5 < \pt < 6$ GeV/c suggests a common partonic basis for the three produced hadrons of diversely different structures.

\begin{figure}[tbph]
\vspace*{-.5cm}
\includegraphics[width=.8\textwidth]{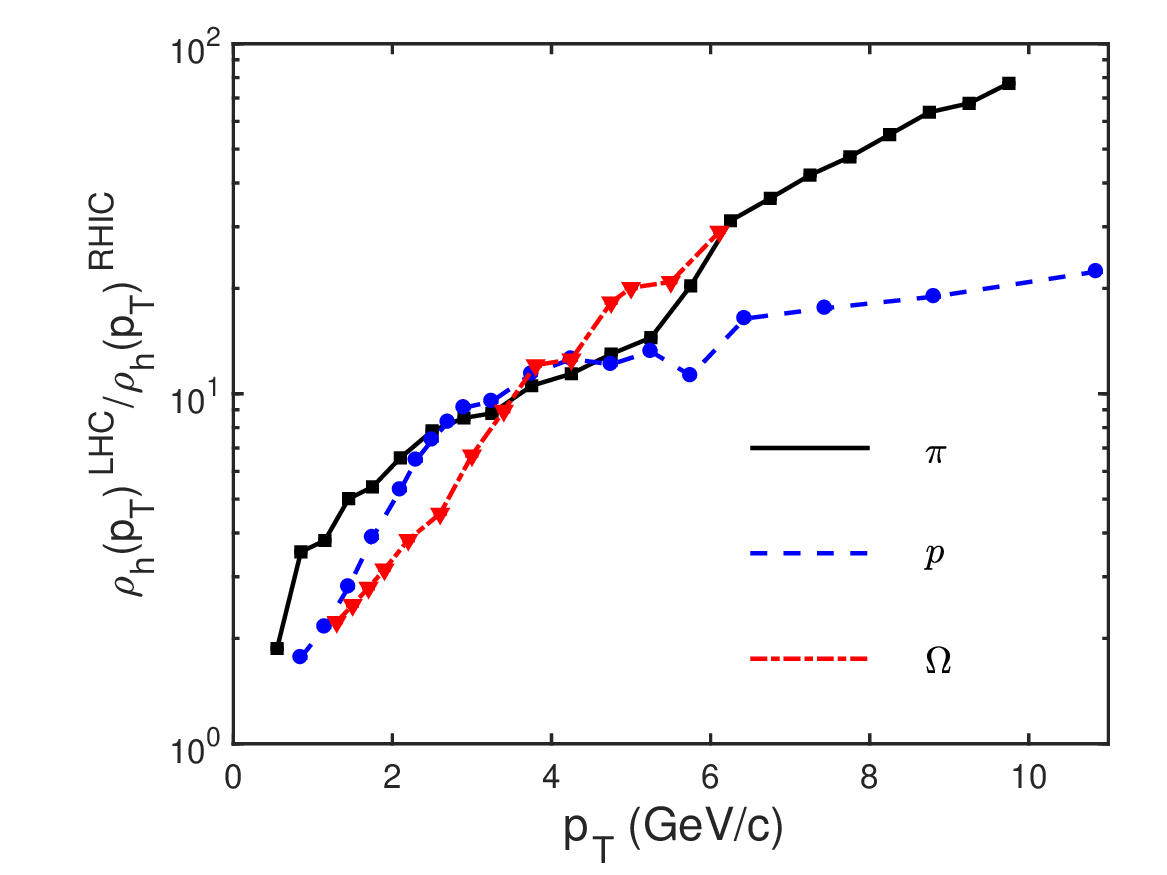}
\caption{(Color online) With $\rho_h(p_T)$ denoting $dN_h/dp_Td\eta|_{\eta\approx 0}$, the ratios $\rho_h^{\rm LHC}(p_T)/\rho_h^{\rm RHIC}(p_T)$ between Pb-Pb collisions at $\sqrt{s_{NN}}=2.76$ TeV and Au-Au collisions at  $\sqrt{s_{NN}}=200$ GeV vs $p_T$ are shown for $h=\pi, p, \Omega$.
 The data are from \cite{Adam:2015kca, ba2, aa, xlz}.}
\end{figure}

Finally, we show another plot of data from LHC that is thought 
provoking. From the $p_T$ distributions of $p$ and $\Omega$
measured by ALICE \cite{Adam:2015kca, ba2}, we plot their ratio vs $p_T$ as shown by the solid (black) line in Fig.\ 2.
The general trend is an exponential rise up to the highest available $p_T$
 with an increase of a factor of 10. The conventional understanding of hadrons produced at
$p_T\sim7$ GeV/c is by the fragmentation of hard scattered gluons
or light-quarks. However, $s$-quark  jets are highly suppressed;
moreover, even if an $s$ quark is produced at high $p_T$, its
fragmentation into $\Omega$ is even more suppressed. To our
knowledge it has never been measured, let alone at $p_T=7$ GeV/c.
Figure 2 shows that the $\Omega/p$ ratio at RHIC in dashed (red) line also grows exponentially until $p_T\approx 3$ GeV/c and then decreases slowly.
The phenomena at both energies are clearly calling for an explanation.

\begin{figure}[tbph]
\vspace*{-0.5cm}
\includegraphics[width=.8\textwidth]{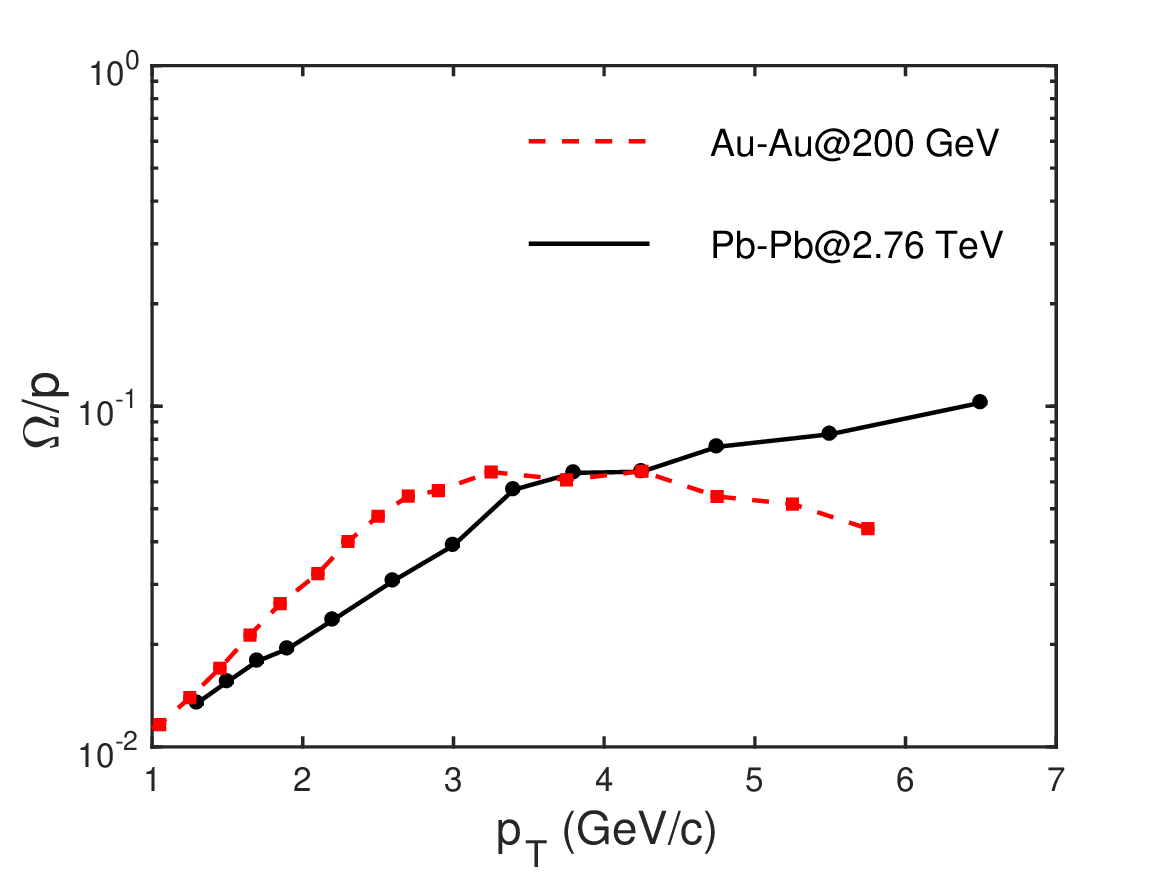}
\caption{(Color online) The ratio $\Omega/p$ vs $p_T$ for central collision in Au-Au collisions at  $\sqrt{s_{NN}}=200$ GeV and Pb-Pb collisions at $\sqrt{s_{NN}}=2.76$ TeV. The data are from \cite{Adam:2015kca, ba2, aa, xlz}.}
\end{figure}

\section{Formulation of the Problem}
To calculate the $p_T$ distribution at mid-rapidity for all
hadrons, we use the same formalism as described earlier for Au-Au
collisions at 200 GeV \cite{hy0, hy1, hz2} and for Pb-Pb collisions at
2.76 TeV \cite{hz1}, i.e., the recombination of thermal and shower
partons. We shall use an improved version of the treatment of
momentum degradation \cite{hy3} and adjust the degradation parameters
to fit the LHC data over a wider range of $p_T$. As a consequence,
the study in Ref.\ \cite{hz1} for $p_T<5$ GeV/c is superseded because
the inclusion of harder jets up to 30 GeV/c with less momentum
degradation results in a profusion of soft shower partons.
Furthermore, we shall include also the production of multi-strange
hadrons. The high density of shower partons introduces another
complication, which is the recombination of partons from different, but adjacent, jets. Although that component turns out not to be dominant, its
effect must be calculated so as to ascertain its magnitude.

The basic framework that describes the recombination of thermal and shower partons at midrapidity 
is intuitively obvious, namely: a convolution of the parton distribution with the recombination function \cite{hy0, hy1, hz1, hz2}
\begin{eqnarray}
p^0{dN^M\over dp_T}&=&\int {dp_1\over p_1}{dp_2\over p_2} F_{q_1\bar q_2}(p_1,p_2) R_{q_1\bar q_2}^M(p_1,p_2,p_T) \label{31} \\
p^0{dN^B\over dp_T}&=&\int \left[\prod_{i=1}^3 {dp_i\over p_i} \right] F_{q_1q_2q_3}(p_1,p_2,p_3) {R}_{q_1q_2q_3}^B(p_1,p_2,p_3,p_T)    \label{32}
\end{eqnarray}
The essence is in the details of what the symbols mean. The LHS of Eqs.\ (\ref{31}) and (\ref{32}) are the invariant $p_T$ distributions of meson and baryon, respectively, averaged over $\eta$ at midrapidity and over all $\phi$. 
They appear as invariant \dis s in the 1D momentum space, but are derived from the invariant \dis\ in 3D as follows:
\begin{eqnarray}
p^0{dN\over dp_T} = p^0p_T{1\over \Delta y}\int_{\Delta y}dy{1\over 2\pi}\int_0^{2\pi}d\phi p^0{d^3N\over d^3p} 
\end{eqnarray}
with $\Delta y$ being a narrow interval at $y\approx 0$, say from $-0.5$ to $+0.5$. Thus our formalism here is not framed to address the global properties of the nuclear collisions, such as total charge multiplicity or long-range correlation.
The parton momenta $p_i$ are the transverse momenta (with the subscript $T$ omitted) of the coalescing quarks. $R^{M}$ and $R^{B}$ are the recombination functions (RFs) for meson and baryon, respectively. The central issue in the formalism is the determination of the parton distribution $F_{q_1\bar{q}_2}$ and $F_{q_1q_2q_3}$ just before hadronization. Because we intend to treat hadron produced in as wide a range in $p_T$ as experimental data on identified particles are available (for pion up to 20 GeV/c), we must consider partons that are produced in soft, semihard and hard scatterings. We group them into two classes of partons, thermal ($\rm T$) and shower ($\rm S$), and use $\cal T$ and $\cal S$ to denote their invariant distributions in $p_i$. Taking into account the recombination of different types of partons, we thus have
\begin{eqnarray}
F_{q_1\bar q_2}&=&{\cal TT+TS+SS}  \label{33} \\
F_{q_1q_2q_3}&=&{\cal TTT+TTS+TSS+SSS}    \label{34}
\end{eqnarray}

The soft partons generated by multiple partonic scattering and radiation in the medium interact with the bulk partons, and cannot be distinguished from the latter by the time the density of all soft partons is low enough for hadronization. They are all referred to here as thermal partons in  the final stage of the quark matter as they move out of the deconfinement phase. 
We emphasize that the adjective {\it thermal} should not be taken to mean that it is to be identified with the thermalized system discussed in hydrodynamical description of an evolved fluid medium. Throughout the expansion phase of the system the bulk medium receives additional contributions of soft partons from the hard and semihard partons that lose energy as they traverse the medium, some of which may never emerge from the medium to form minijets or high-momentum jets. We use {\it thermal} mainly to distinguish those partons from the shower partons, and partly to retain the same nomenclature as used before.
The  shower partons that we consider are the fragmentation products of the hard and semihard partons that emerge from the surface after momentum degradation.
The usual fragmentation function describes the distribution of hadrons in a hard-parton jet, but in our view there is the intermediate stage of shower partons which recombine among themselves to form hadrons \cite{hy4}.  In heavy-ion collisions those shower partons are in the environment of  
thermal partons and can therefore undergo TS recombination. Those are the $\cal T$ and $\cal S$ distributions in Eqs.\ (\ref{33}) and (\ref{34}). While that possibility has always been an essential part of the RM, our point here is that at LHC the thermal partons include the soft partons generated by the hard and semihard partons dissipated in the expanding medium.

We use a simple exponential form to represent the thermal parton distribution (which is another reason to call it {\it thermal})
\begin{eqnarray}
{\cal T}(p_1) = p_1{ dN^T_q\over dp_1} = Cp_1e^{-p_1/T}    \label{35}
\end{eqnarray}
with the  prefactor $Cp_1$ necessary to yield pure exponential behavior for the pion distribution $dN^{\pi}/p_Tdp_T\propto C^2\exp(-p_T/T)$ arising from $\rm TT$ recombination only, as observed at low \ppt. Of course, by inspection $C$ has the dimension of inverse momentum. $T$ is the inverse slope that should not be given the interpretation of temperature in the sense considered in conventional hydrodynamical models.  
For our use at LHC now, we need to make sure that the form in Eq.\ (\ref{35}) is consistent with the pure exponential behavior of the baryon function $B_h(s,N_{\rm part},\pt)$ that we have observed in \cite{Hwa:2018qss}, where it is defined for baryon $h$
\bq
B_h(\pt)={m_T^h\over \pt^2} {d{\bar N}_h\over \pt d\pt}(\pt) ,  \label{3.6}
\eq
with $m_T^h=(m_h^2+p_T^2)^{1/2}$. 
Identifying $d{\bar N}_h/d\pt$ in the above equation with $dN^B/d\pt$ in Eq.\ (\ref{32}), and using only the ${\cal TTT}$ term on the right-hand side of (\ref{34}) with ${\cal T}$ being given in (\ref{35}), we use the simple form for the RF of $\Omega$
\bq
R^\Omega_{sss}(p_1,p_2,p_3,\pt) \propto  {\prod_{i=1}^3} \delta ( p_i/ \pt - 1/3 ) ,  \label{3.7}
\eq
and obtain
\bq
B_\Omega(\pt) \propto {m_T^\Omega\over p^0} C^3 \exp (-\pt/T) .  \label{3.7}
\eq
At mid-rapidity considered here and in \cite{Hwa:2018qss}, we have $p^0\approx m_T$, so the thermal contribution to Eq.\ (\ref{3.6}) based on (\ref{35}) indeed leads to pure exponential behavior. By similar considerations $B_h(\pt)$ for other baryons $h$ will also have exponential dependence on \ppt\ as a consequence of (\ref{35}), as we shall show explicitly in the following sections. 

The value of the inverse slope $T$ depends on the strangeness content of $h$, as well as $\sqrt{s_{NN}}$. For $h=p$ and $\Omega$ at $\sqrt{s_{NN}}=2.76$ TeV, we found in \cite{Hwa:2018qss}
\bq
T_p=T_q=0.39\ {\rm GeV/}c ,		\label{3.8}  \\
T_\Omega=T_s=0.51\ {\rm GeV/}c ,		\label{3.9}
\eq
where the subscripts $q, s$ denote light and strange quarks. Those are values deduced from pure phenomenology in \cite{Hwa:2018qss} and are far from any temperature discussed in hydrodynamical treatment of the hot dense matter as a fluid.

Returning to the general discussion about Eqs.\ (\ref{33}) and (\ref{34}), when shower partons are important at low $p_T$, then $\cal TS, TTS$ and $\cal TSS$ components need to be included. Nevertheless, we retain the form of $\mathcal T(p_1)$ in Eq.\ (\ref{35}) for the thermal component.

The shower parton distribution after integration over jet momentum $	q$ and summed over all jets is 
\begin{eqnarray}
{\cal S}^j(p_2)=\int {dq\over q}\sum_i \hat F_i(q) S_i^j(p_2, q),  \label{36}
\end{eqnarray}
where $\hat F_i(q)$ is the distribution of hard or semihard parton of type $i$ at the medium surface after momentum degradation while transversing the medium but before fragmentation. $\hat F_i(q)$ was introduced previously for collisions at RHIC for any centrality \cite{hy3, hz3}, but will be modified below to suit our description of the physics at LHC. $S_i^j(z)$ is the unintegrated shower-parton distribution (SPD) in a jet of type $i$ fragmentation into a parton of type $j$ with momentum fraction $z$. It is determined from the fragmentation function (FF) on the basis that hadrons in a jet are formed by recombination of the shower partons in the jet \cite{hy4, hy5}. In particular, the recombination of a quark $j$ with an antiquark $\bar j$ in a jet of type $i$ forms a pion, for which the FF is $D_i^{\pi}(z_j+z_{\bar j})$. The numerical form for $S_i^j(z_j)$ can therefore be calculated from the data on $D_i^{\pi}$ and the RF for pion. Note that $S_i^j(z_j)$ is unaffected by nuclear medium because it describes the shower partons that are the fragmentation products of hard and semihard partons outside the medium.

The RFs were introduced a long time ago \cite{dh, rh1}  and have been applied successfully to many collision processes \cite{hy6, hy7, hy1, hy2, hz1, hz3}. Here for brevity we give only the RFs for pion and proton, leaving other hadrons to be specified later as the cases arise,
\begin{eqnarray}
R^{\pi}_{q\bar q}(p_1, p_2, p_T)&=&\frac{p_1p_2}{p_T}\delta(p_1+p_2-p_T),  \label{37} \\
R^{p}_{uud}(p_1, p_2, p_3, p_T)&=&g_{st}^pg_p(y_1y_2)^{\alpha}y_3^{\beta}\delta(\sum\limits_iy_i-1),\qquad y_i=\frac{p_i}{p_T} ,   \label{38}
\end{eqnarray}
where $g_{st}^p=1/6$, $\alpha=1.75$, $\beta=1.05$, and
\begin{eqnarray}
g_p=[B(\alpha+1, \alpha+\beta+2)B(\alpha+1, \beta+1)]^{-1}, \label{39}
\end{eqnarray}
$B(a, b)$ being the Beta function. 

As a note of affirmation, we recall that with these RFs used in Eqs.\ (\ref{31}) and (\ref{32}), and considering only the $\cal{TT}$ ($\cal{TTT}$) component for pion (proton), we have been able to fit the pion and proton spectra for $1<p_T<2$ GeV/c in Au-Au collisions at 200 GeV \cite{ssa} with a common value of the inverse slope in Eq.\ (\ref{35}) \cite{hz3}.
For $p_T<1$ GeV/c there is resonance contribution that Eq.\ (\ref{31}) does not account for, while for $p_T>2$ GeV/c shower parton contributions invalidate the approximation of $F_{q\bar q}$ and $F_{uud}$ by $\cal {TT}$ and $\cal{TTT}$, respectively. In the $1<p_T<2$ GeV/c interval one may find the excellent agreement with data surprising, when only the exponential form of Eq.\ (\ref{35}) is used for both pion and proton, since the proton data for $dN^p/p_Tdp_T$ is not exponential. However, it is precisely because of the momentum dependence in $R^p$ in Eq.\ (\ref{38}) and the fact that $p_0$ in Eq.\ (\ref{32}) is the transverse mass $m_T(p_T)$ at $y=0$ that renders $dN^p/p_Tdp_T$ to deviate from pure exponential. The phenomenological success there gives strong support to the recombination model. 
Thus the essence of this work is to calculate the effects of the shower partons in the  intermediate $p_T$ region in collisions at LHC.

\section{Shower Parton Distributions}
Focusing on the shower partons, we see in Eq.\ (\ref{36}) that $\hat F_i(q)$ is the distribution to be determined for collisions at LHC, since $S_i^j(p_2, q)$ is the SPD outside the nuclear medium and is independent of the collision system; it has been determined previously 
from FFs in vacuum \cite{hy4}. At any particular 
impact parameter $b$, $\hat F _i(q, b)$ is the average over azimuthal angle $\phi$ of $\bar F_i(q, \phi, b)$, which has three essential parts \cite{hy3}
\begin{eqnarray}
\bar F_i(q, \phi, b)=\int d\xi P_i(\xi, \phi, b)\int dkkf_i(k)G(k, q, \xi), \label{41}
\end{eqnarray}
where $f_i(k)$ is the parton density in the phase space $kdk$ at the point of creation, $k$ being the initial momentum of the hard or semihard parton $i$, and $P_i(\xi, \phi, b)$ is the probability for the parton $i$ to have a dynamical path length $\xi$ at $\phi$ and $b$. The two parts are connected by $G(k, q, \xi)$
\begin{eqnarray}
G(k, q, \xi)=q\delta(q-ke^{-\xi}), \label{42}
\end{eqnarray}
which is the momentum degradation function, relating the initial parton momentum $k$ to the final momentum $q$ at the medium surface by an exponential decay in $\xi$, the length that carries all the geometrical and dynamical information of the process through $P_i(\xi, \phi, b)$. The details of calculating  $P_i(\xi, \phi, b)$ are given in Ref.\ \cite{hy3} and summarized in the Appendices in Ref.\ \cite{hz2}. We shall recall the essence below in order to re-parametrize it for suitable use at LHC.

First, we need to state why we describe momentum degradation in the way outlined above without adopting the results obtained by pQCD in the literature. Because we intend to calculate the $p_T$ \dis s of all hadrons from 1 to 20 GeV/c, we need to let $q$ in Eq.\ (\ref{36}) be integrated from low values in order for the shower partons to have their momenta be as low as 0.5 GeV/c. In practice, $q$ is integrated from 2 to 30 GeV/c. Low-order perturbative QCD is not reliable for virtuality less than 8 GeV/c, so the major portion of the contribution to the shower partons in the soft region cannot make use of the established theory. Furthermore, the usual calculation based on DGLAP evolution equation is on medium modification of the fragmentation function, while we need shower parton \dis\ for the purpose of recombination. The dependence on the medium is usually described in terms of entropy density and local flow velocity, which are hydrodynamical quantities tuned to fit low-$p_T$ data, which are exactly what we attempt to reproduce in addition to intermediate-$p_T$ data independent of fluid dynamics. For these reasons we use a phenomenological procedure that has been shown to generate the azimuthal and $p_T$ dependencies of $R_{AA}(\phi,p_T)$ at RHIC \cite{hy3} and can readily be extended to higher energy, as we now proceed to do.

The initial momentum distributions have been determined in Ref.\ \cite{sgf} for Au-Au collisions at 200 GeV and Pb-Pb collisions at 5.5 TeV. They are parametrized in the form
\begin{eqnarray}
f_i(k)=K\frac{A}{(1+k/B)^{\beta}}. \label{43}
\end{eqnarray}
We make logarithmic interpolations of the parameters between the two energies for $\ln A$, $B$ and $\beta$ and obtain for $\sqrt{s_{NN}}=2.76$ TeV the parameters shown in Table I with $K=2.5$.

\begin{table}
\tabcolsep0.2in
\begin{tabular}{|c|c|c|c|c|c|c|}
\hline
 & $g$ &  $u$ &  $d$ & $\bar u$ &  $\bar d$  &s, $\bar s$\\ 
 \hline
 $A$ [$10^4$/GeV$^2$] & 6.2 &1.138 &1.266 &0.24 &0.23& 0.093\\
 $B$ [GeV]& 0.98 &0.687 &0.677 &0.87 &0.88 &1.05\\
 $\beta$ & 6.22 &5.67 &5.66 &5.97 &5.99 &6.12\\
 \hline
 \end{tabular}
 \caption{Parameters for $f_i(k)$ in Eq.\ (\ref{43}).} \label{table1}
 \end{table}

The connection between geometry and dynamics is imbedded in the probability function $P_i(\xi,\phi,b)$. 
The geometrical path length $\ell$,  when written more fully, is
\begin{eqnarray}
\ell(x_0, y_0, \phi, b)=\int_0^{t_1(x_1, y_1)}dt D(x(t), y(t)) \label{44}
\end{eqnarray}
that is calculable from nucleon geometry. The transverse coordinate $(x_0, y_0)$ is the initial point of creation of a hard parton, and $(x_1, y_1)$ is the exit point. The integration is weighted by the local density, $D(x, y)$, along the trajectory, which is marked by the variable $t$ that does not denote time. As the medium expands, the end point $t_1(x_1, y_1)$ increases, but $D(x(t), y(t))$ decreases, so $\ell$ is insensitive to the details of expansion dynamics. The dynamical path length $\xi$ is proportional to $\ell$, but is to be averaged over all initial points $(x_0, y_0)$, i.e.,
\begin{eqnarray}
P_i(\xi, \phi, b)=\int dx_0dy_0Q(x_0, y_0, b)\delta(\xi-\gamma_i\ell(x_0, y_0, \phi, b)) \label{45}
\end{eqnarray}
where $Q(x_0, y_0, b)$ is the probability that a hard (or semihard) parton is produced at $(x_0, y_0)$, calculable from nucleon thickness functions \cite{hy3, hz2}. The only parameter that we cannot calculate is $\gamma_i$, which incorporate the effects of energy loss during the passage of the parton through the non-uniform and expanding medium. The average dynamical path length $\bar\xi_i$, defined by
\begin{eqnarray}
\bar\xi_i(\phi, b)=\int d\xi\xi P(\xi, \phi, b), \label{46}
\end{eqnarray}
depends on geometry, and is proportional to $\gamma_i$, as can readily be seen upon substituting Eq.\ (\ref{45}) into (\ref{46}). Thus, using Eqs.\ (\ref{41})-(\ref{45}), $\hat F_i(q, b)$ can be calculated once $\gamma_i$ are specified.

In treating hadron production at RHIC we have determined $\gamma_i$ in Ref.\ \cite{hz2} and obtained excellent fits of the $p_T$ distributions of $\pi, K, p$ for $p_T<10$ GeV/c at all centralities \cite{ph1, ph2, ph3, st1, ph4, st2}. We used $\gamma_g=0.14$ for gluon and $\gamma_q=0.07$ for all light quarks, their ratio being 2 as an approximation of the color factor $C_A/C_F=9/4$. Because $\bar\xi_i(\phi, b)\propto\gamma_i$, we have $\bar\xi_g(\phi, b)/\bar\xi_q(\phi, b)=2$, which directly implies that gluons on average lose the same fraction of momentum as quarks do in half  the distance of traversal through the nucleon medium. That turned out to be an important factor in enabling us to reproduce both the pion and proton spectra because at intermediate $p_T$ pions are more affected by semihard gluon minijets, while protons are more so by quark minijets, due to their recombination characteristics \cite{hz2}.

To extend the treatment of momentum degradation to collisions at LHC, we cannot expect $\gamma_i$ to remain the same as at RHIC. It has been found that the nuclear modification factor $R_{AA}$ for Pb-Pb collisions at 2.76 TeV at 0-5\% centrality decreases rapidly from $p_T=2$ GeV/c to a minimum value of 0.13 at $p_T=$ 6-7 GeV/c, after which there is a significant rise, reaching $R_{AA}\approx 0.4$ for $p_T>30$ GeV/c \cite{raa}. Such data suggest that jet quenching becomes less severe at higher momentum, so $\gamma_i$ should decrease as the hard parton momentum increases. Hence, we parametrize $\gamma_g$ as
\begin{eqnarray}
\gamma_g(q)=\frac{\gamma_0}{1+(q/q_0)^2}, \label{47}
\end{eqnarray}
with $\gamma_0$ and $q_0$ to be determined by fitting the hadronic spectra in the intermediate $p_T$ region, and we continue to set $\gamma_q=\gamma_g/2$ as before. Although the $p_T$ distributions will not be computed until Sec.\ VI after several other issues are discussed, we give here the values $\gamma_0=2.8$ and $q_0=7$ GeV/c that will be determined there, so that our present discussion can proceed with concrete numerical specificity to show the nature of physics involved. Furthermore, we shall hereafter be concerned with only the most central collisions 0-5\%. We shall therefore omit the symbol $b$ and perform all calculation with the appropriate range of impact parameter. Defining $\hat F_i(q)$ as the average of $\bar F_i(q, \phi)$ over $\phi$
\begin{eqnarray}
\hat F_i(q)=\frac{1}{2\pi}\int_0^{2\pi}d\phi\bar F_i(q, \phi), \label{48}
\end{eqnarray}
we can, using Eqs.\ (\ref{42}-\ref{45}) and following the details discussed in Refs.\ \cite{hy2, hz2}, compute $\hat F_i(q)$ for all parton types $i$ listed in Table I, and for all $q<30$ GeV/c. Although the hadron transverse momentum $p_T$ will not exceed 20 GeV/c in our calculation, so that $p_2$ in Eqs.\ (\ref{31}) and (\ref{32}) is also less than that upper limit, it is necessary to consider higher values of $q$ because of the integration in Eq.\ (\ref{36}).
In Fig.\ 3(a) we show $\hat F_g$ for gluon by the solid line, and in (b) $\hat F_i$ for $i=q$, $\bar q$ and $s$ by other line types, assuming that $\gamma_s=\gamma_q$. 
They are compared to $q^2f_{g, q}(q)$ for no momentum degradation (i.e., $\xi=0$) shown by the lines of open symbols. We recall that $f_i(k)$ is the initial parton distribution defined in the phase space $kdk$, while $\hat F_i(q)$ is the invariant distribution in $dq/q$. It is possible to see from Fig.\ 3 that the ratio $\hat F_i(q)/q^2f_i(q)$ increases with increasing $q$. That is a consequence of $\gamma_g(q)$ decreasing with $q$, as indicated in Eq.\ (\ref{47}).

\begin{figure}[tbph]
\vspace*{-0.5cm}
\includegraphics[width=.8\textwidth]{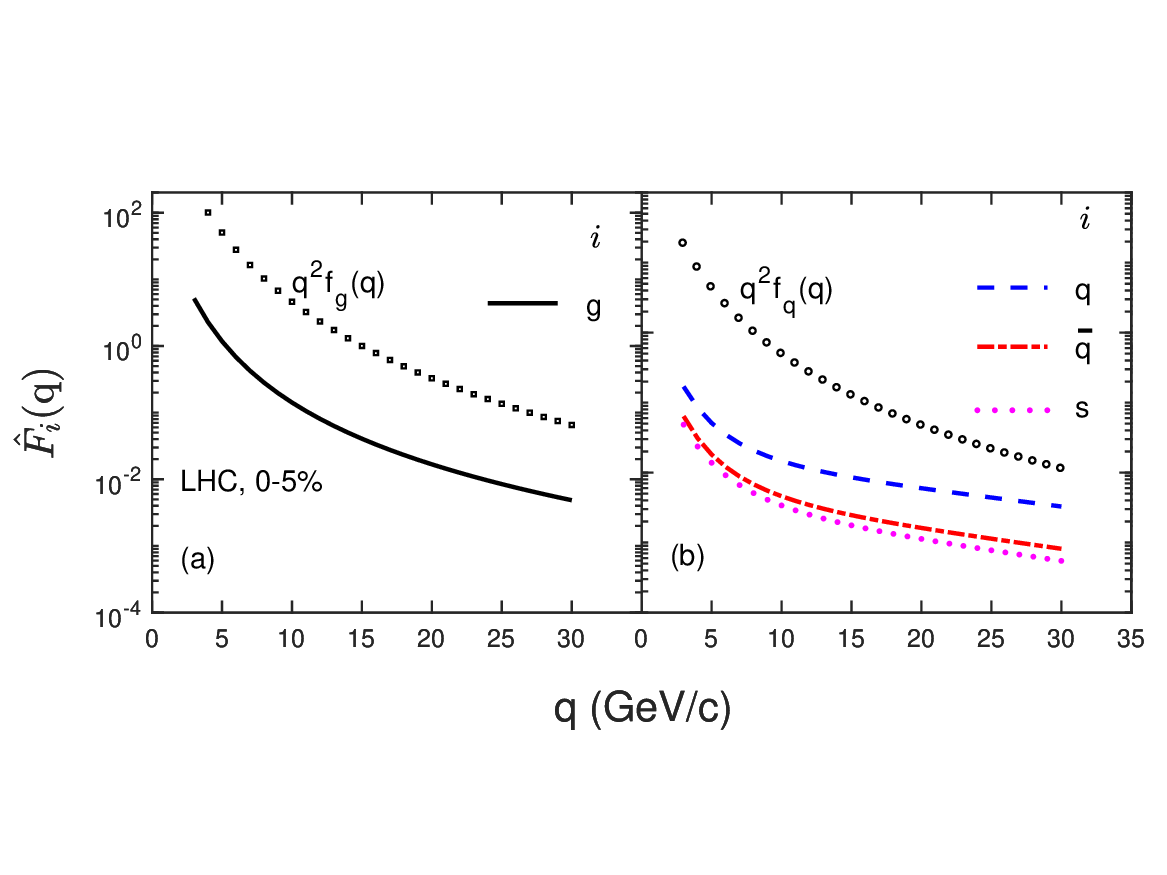}
\vspace*{-1.4cm}
\caption{(Color online) Distribution of minijets at medium surface for 0-5\% centrality. Index $i$ denotes the parton type: (a) $i=g$ for gluon, (b) $i=q$, $\bar q$, $s$ (with $\bar s$ being treated the same as $s$). The line with open squares in (a) represents the distribution of gluons without momentum degradation; the line with open circles in (b) represents the same for light quarks.}
\end{figure}

With $\hat F_i(q)$ now known explicitly, we can proceed to the calculation of $\mathcal{S}^j(p_2)$ in Eq.\ (\ref{36}). The SPDs $S_i^j(p_2, q)$ are derived in Refs.\ \cite{hy4} and summarized in \cite{hz2}. Since the fragmentation of hard and semihard partons into shower partons takes place outside the medium in our treatment, the structure of SPDs is independent of the collision energy. Thus $\mathcal{S}^j(p_2)$ at LHC differs from that at RHIC only because $\hat F_i(q)$ is now enhanced, not because of any changes in $S_i^j(p_2, q)$. While $i$ in Eq.\ (\ref{36}) is summed over all parton types listed in Table I, $j$ will only be $u$, $d$, $s$ and their antiquarks because in our formalism of recombination gluons do not directly participate in hadronization. They are always converted to $q\bar q$ pairs first, which dress themselves before becoming the constituent quarks of the produced hadrons \cite{rh1}. The conversion of gluons to $q\bar q$ pairs are referred to as enhancing the sea for hadronization at large rapidity \cite{rh1, hy7}. Here at large $p_T$ the same concept of gluon conversion applies, except that instead of enhancing the sea each $q$ and $\bar q$ can participate in forming a hadron, but in single-particle inclusive distribution only the leading partons with large momentum fractions are considered in the calculation.

Before showing the result from calculating $\mathcal{S}^j(p_2)$, we note that in using Eq.\ (\ref{36}) in practice, apart from  $q$ being integrated from $q=2$ to 30 GeV/c, as mentioned earlier,  the SPD $S_i^j(p_2, q)$ is made to deviate from the scaling form $S_i^j(z)$ by our insertion of a cutoff factor $c_2(p_2)$
\begin{eqnarray}
S_i^j(p_2, q)=S_i^j(p_2/q)c_2(p_2), \label{410}
\end{eqnarray}
 where
 \begin{eqnarray}
c_2(p_2)=1-e^{-(p_2/p_c)^2}, \hspace{0.5cm} p_c=0.5 \mbox{GeV/c}. \label{411}
\end{eqnarray}
 Such a factor is necessary to render the shower partons meaningful in the soft region, for otherwise the IR divergent FF, $D_i(p_T/q)$, as $p_T\to 0$, would lead to unrealistically large $S_i^j(p_2/q)$. This point is discussed in Appendix C of Ref.\ \cite{hz2}, where $c_2(p_2)$ is denoted by $\gamma_2(p_2)$. The value of $p_c$ in Eq.\ (\ref{411}) is chosen so that we can obtain a good fit of the proton spectrum at low $p_T$, as will be shown in Sec.\ VI. By relinquishing our claim for any reliability of our model predictions in the region $p_T<1$ GeV/c, we find that what we can calculate at $p_T>1$ GeV/c is insensitive to the precise value of $p_c$. We use $p_c=0.5$ GeV/c just to fit the proton spectrum at $p_T<1$ GeV/c. 
 Note that we use the proton distribution as the guide, not pion, because there are resonance and other contributions to the pion distribution at very low $p_T$. The details will become more clear when the mathematical expressions for recombination are shown explicitly below.
 
  \begin{figure}[tbph]
\includegraphics[width=.8\textwidth]{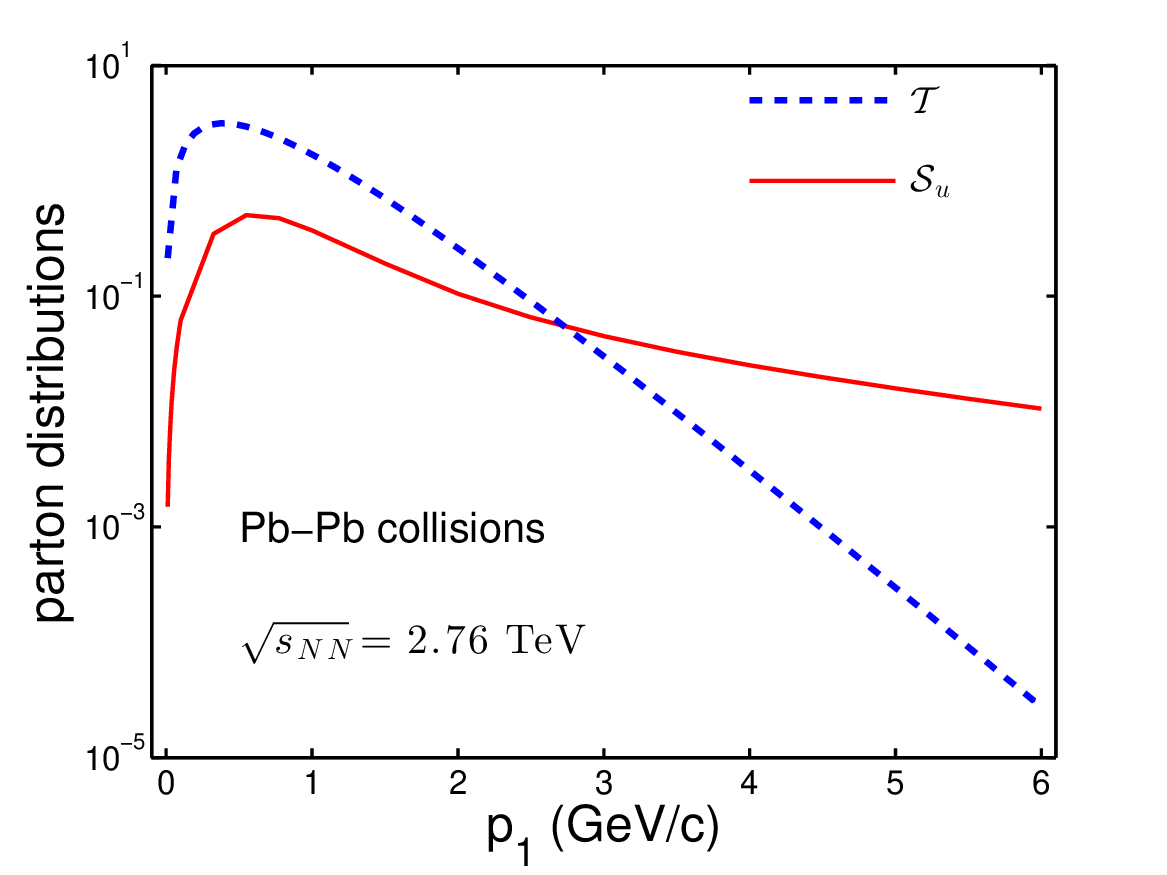}
\caption{(Color online) Thermal distribution $\mathcal{T}(p_1)$ is depicted by the  dashed (blue) line for $T=0.39$ GeV. Shower parton distribution $\mathcal{S}^u$ is shown in  solid (red) line with  low-$p_1$ cutoff.} 
\end{figure}

 Substituting Eqs.\ (\ref{410}) and (\ref{411}) into (\ref{36}), we obtain the invariant shower-parton distribution $\mathcal S^j(p_2)$ after integrating over $q$ and summing over all initiating partons $i$. For $j=u$, it is shown in Fig.\ 4 by the solid (red) line, plotted against $p_2$ but labeled as $p_1$, since it is to be compared to the thermal parton distribution $\mathcal T(p_1)$ in the same figure. For $\mathcal T(p_1)$ we use Eqs.\ (\ref{35})  and (\ref{3.8}) with the value of $C$ to be discussed in Sec.\ VI. 
 The $\mathcal T(p_1)$ distribution is shown by the  dashed (blue) line in Fig.\ 4. Evidently, $\mathcal S(p_1)$ dominates over $\mathcal T(p_1)$ for all $p_1>3$ GeV/c.  Hereafter, for the sake of brevity we omit the superscript of quark type $j$ in $\mathcal S^j(p_1)$, as we routinely do for $\mathcal T(p_1)$, when no confusion is likely to ensue.

\section{Two-jet Recombination}
Before we embark on the actual task of computing the inclusive distributions, we discuss an issue that should arise upon examining Fig.\ 4. Eqs.\ (\ref{33}) and (\ref{34}) display only the schematic structure of the various components, and are adequate only as a general layout for use in Eqs.\ (\ref{31}) and (\ref{32}). Kinematic constraints on the shower-parton momenta that will be shown in detail in the next section result in the contribution from $\mathcal{SS}$ and $\mathcal{SSS}$ terms to be dominant only in the large $p_T$ region. There is another type of shower-parton recombination that has not been discussed above; that is the subject of our consideration in this section.

In Refs.\ \cite{hy1, hz1, hz2} where $\rm SS$ recombination is considered, the shower partons arise from the same jet. (The same applies to  $\rm SSS$ for baryons as well, but will not be reiterated.) Such a term is equivalent to fragmentation, since it is from the FF, $D_i^{\pi}(z)$, that the SPDs are derived in the first place \cite{hy4}. In view of the dominance of $\mathcal{S}(p_1)$ over $\mathcal{T}(p_1)$ for $p_1>3$ GeV/c, it is reasonable to expect the integral of $\mathcal{S}(p_1)\mathcal{S}(p_2)$ to be larger than $\mathcal{T}(p_1)\mathcal{S}(p_2)$ when convoluted with the same RF, $R^{\pi}(p_1, p_2, p_T)$. At this point it is important for us to be more explicit with indices and distinguish one-jet and two-jet recombinations, which we shall denote by $\rm (SS)^{1j}$ and $\rm (SS)^{2j}$, respectively.

In Fig.\ 5 we show the diagrams in the transverse plane for three types of recombination: (a)
$\rm TS$, (b) $\rm (SS)^{1j}$ and (c) $\rm (SS)^{2j}$. In the notation of Eq.\ (\ref{42}), $k$ is the momentum of the hard or semihard parton at creation, and $q$ is the momentum at the medium surface. The thick red vectors have the dual role of representing the jet momentum in the medium and the degradation effect described by $G(k, q, \xi)$. The thinner red lines outside the medium are the semihard partons $q_j$, which can emit shower partons represented by the thinnest red lines denoted by $p_j$.  The blue dashed arrows are thermal partons.  Recombination is represented by a large black blob with the outgoing open arrow depicting the produced pion.
We emphasize that the shower parton lines are inclusive in the sense that only the ones contributing to the formation of the observed hadron are shown. In particular, a gluon generates a cluster of partons which cannot all be depicted. Thus quark types and baryon numbers cannot be recognized from the schematic diagrams. Furthermore, the
lengths and angles of the vectors are not drawn to scale due to the limitation in presenting the figures clearly, and should not be taken literally.

\begin{figure}[tbph]
\centering
\vspace*{-9cm}
\includegraphics[width=0.9\textwidth]{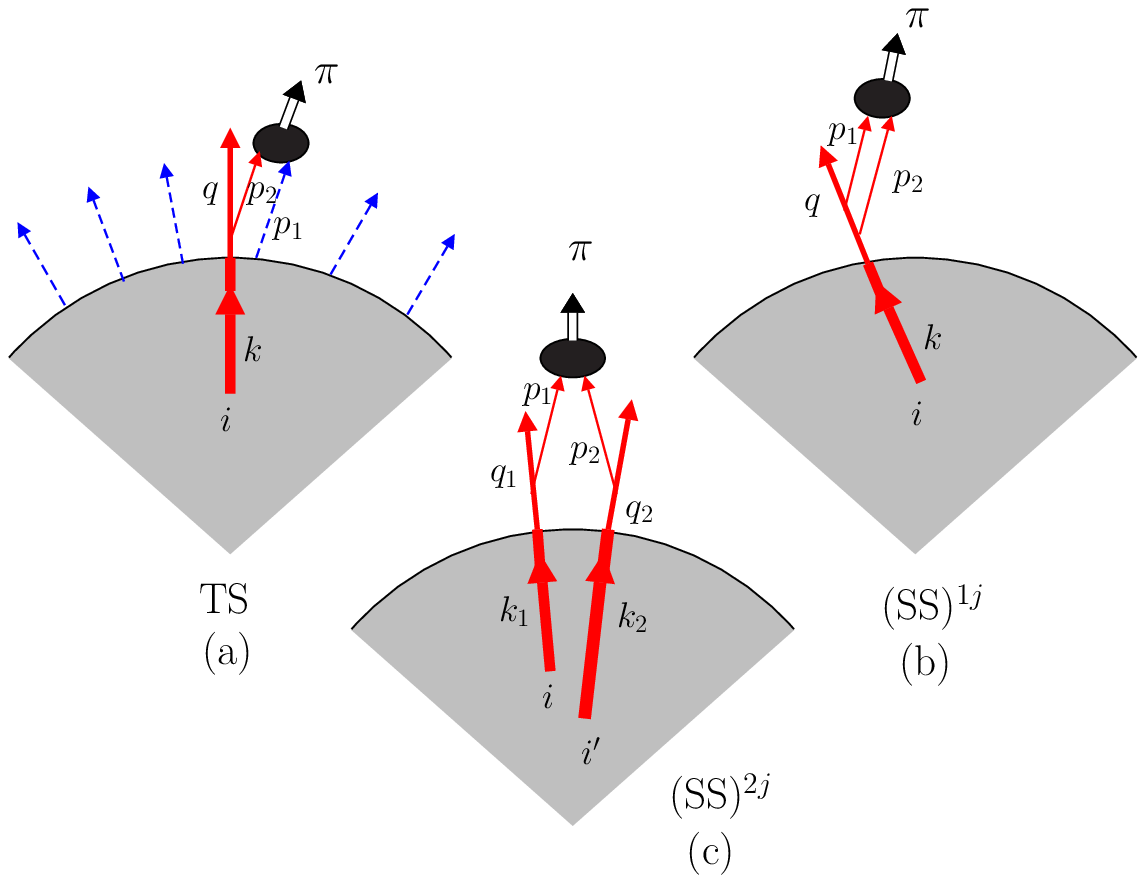}
\vspace*{-.5cm}
\caption{(Color online) Schematic diagrams for parton recombination of (a) TS, (b) SS in one jet, and (c) SS in two jets. Thick (red) lines represent partons in medium, thin (red) lines partons out of medium, thinnest (red) lines shower partons, and dashed (blue) lines thermal partons.
All lines are inclusive in the sense described in the text.}
\end{figure}

Note that in Fig.\ 5(a) and (b) the hard or semihard partons are labeled by $i$, while in (c) the two partons are labeled by $i$ and $i'$. Therein lies the essential point that $\rm TS$ and $\rm{(SS)^{1j}}$ each involves only one jet of type $i$, while $\rm (SS)^{2j}$ involves two jets of types $i$ and $i'$. Thus for $\rm{TS}$ and $\rm{(SS)^{1j}}$ there is only one hard scattering contained in $\hat F_i(q)$, while for $\rm{(SS)^{2j}}$ there are two hard scatterings contained separately in $\hat F_i(q_1)\hat F_{i'}(q_2)$. More explicitly, but leaving out integration over $q$ and summation over $i$ for now (with full expression to be shown in the next section), we have
\begin{eqnarray}
\hat F_i(q)\widehat{TS}(q, p_T)&=&\int\frac{dp_1}{p_1}\frac{dp_2}{p_2}\hat F_i(q)\mathcal {T}^{\bar q}(p_1) {S}_i^q(p_2, q)R_{q\bar q}^{\pi}(p_1, p_2, p_T), \label{51}  \\
\hat F_i(q)\widehat{SS}(q, p_T)&=&\int\frac{dp_1}{p_1}\frac{dp_2}{p_2}\hat F_i(q)\left\{{S}_i^{q}(p_1, q),{S}_i^{\bar q}(p_2, q)\right\}  R_{q\bar q}^{\pi}(p_1, p_2, p_T)\\ \nonumber
&=&\hat F_i(q)\frac{p_T}{q}D_i^{\pi}(p_T, q), \label{52}
\end{eqnarray}
while for $(\rm{SS})^{2j}$ we need to retain the $\phi$ variable in $\bar F_i(q, \phi)$ before it is averaged over $\phi$ in Eq.\ (\ref{48}):
\begin{eqnarray}
\widehat{\cal SS}^{2j}=\int\left[\prod\limits_{a=1}^2\frac{dp_a}{p_a}d\phi_a\right] \bar F_i(q_1, \phi_1)\bar F_{i'}(q_2, \phi_2){{S}_i^{q}(p_1, q_1)\rm {S}_{i'}^{\bar q}(p_2, q_2)}{\bf R}_{\Gamma}^{\pi}(p_1, \phi_1, p_2, \phi_2, p_T, \phi). \label{53}
\end{eqnarray}
Because there are two initiating hard partons $i$ and $i'$ we need to integrate over their respective azimuthal angels $\phi_1$ and $\phi_2$, allowing the RF $\bf R_{\Gamma}^{\pi}$
 to play the role of restricting $\phi_1$ and $\phi_2$ to be nearly equal for the coalescence process to take place. Non-parallel partons have large relative momentum transverse to $\vec{p}_1+\vec{p}_2$, which should not exceed the binding energy of the constituents of the hadron that it is to be formed. That is different from large relative longitudinal momentum parallel to $\vec p_1+\vec p_2$ because in the parton model the momentum fractions of partons in a hadron can vary from 0 to 1 . 
 
 The azimuthal angles $\phi_1$ and $\phi_2$ may be given by a Gaussian distribution in $|\phi_1-\phi_2|$ with an appropriate width. However, since $\phi_1$ and $\phi_2$ are integrated over in Eq.\ (\ref{53}), it is simpler to adopt a factorizable form that requires the partons to be parallel but with a suitable
normalization factor $\Gamma$ that we can estimate, i.e.,
\begin{eqnarray}
{\bf R}_\Gamma^\pi(p_1,\phi_1,p_2,\phi_2,p_T,\phi) = \Gamma\delta(\phi_1-\phi_2)\delta\left({\phi_1+\phi_2\over 2} - \phi\right) R^\pi(p_1,p_2,p_T),
\label{54}
\end{eqnarray}
where $\Gamma$ is the probability that two parallel partons can recombine.  Since the partons are emitted from the medium at early times, we may
consider the emitting system as being a thin almond-shaped overlap region viewed from its side in the same transverse plane at midrapidity as where
the pion is detected.  For 0-5\% centrality the almond is almost circular.  The partons at $\phi_i$ are parallel, but can be emitted at any
distance from the center of the circle.  Looking at the emitting source edgewise, it is essentially a one-dimensional system of width approximately 10
fm, which is slightly less than $2R_A$ since high-density partons are not likely to be emitted tangentially from the edges.  The two parallel partons
should be separated by a distance not greater than the diameter of a pion ($\sim 1$ fm), given that the jets have some width.  Thus our estimate for
$\Gamma$ is the ratio $\sim 1/10$.  We do not see that any more elaborate analysis of the coalescence process can provide a more transparent
description of ${\bf R}_\Gamma^\pi$. Applying Eq.\ (24) to (23) we obtain upon averaging over $\phi$
\begin{eqnarray}
\widehat{\cal SS}^{2j}=\Gamma\int\frac{dp_1}{p_1}\frac{dp_2}{p_2}\hat F_i(q_1) {S}_i^{q}(p_1, q_1)\hat F_{i'}(q_2){{S}_{i'}^{\bar q}(p_2, q_2)}R^{\pi}(p_1, p_2, p_T). \label{55}
\end{eqnarray}
By comparing this equation with Eq.\ (22) we see that the 2j contribution has an extra factor of $\Gamma\hat F_i(q_2)$ with $p_2$ ranging from 0 to $q_2$. On the other hand, the symmetrization of the two shower-parton product in the 1j contribution, when expressed in terms of momentum fractions $x_i=p_i/q$, reveals the ranges $0<x_2<1-x_1$, and $0<x_1<1-x_2$ in the two terms
\begin{eqnarray}
\{ {S}_i(x_1), {S}_i(x_2)\}=\frac{1}{2}\left[{S}_i(x_1) S_i(\frac{x_2}{1-x_1})+S_i(x_2) S_i(\frac{x_1}{1-x_2})\right]. \label{56}
\end{eqnarray}
Thus, when two shower partons are in the same jet, the sum of their momenta, $p_1+p_2$, cannot exceed the jet momentum $q$. That is the kinematical restriction mentioned in the beginning of this section, and corresponds to the familiar condition that $p_T<q$ in the FF $D_i^{\pi}(p_T, q)$ in Eq.\ (22).

Since the large-$q$ dependence of $\hat F_i(q)$ is power-law behaved, the $\rm{(SS)}^{1j}$ component dominates at high $p_T$, where the components involving the thermal partons (i.e. $\rm TT$ and $\rm {TS}$) are damped due to the exponential behavior of $\mathcal{T}(p_1)$. The $\rm (SS)^{2j}$ component involves $\hat F_i(q_1)$ and $\hat F_{i'}(q_2)$ in Eq.\ (\ref{55}) so it is suppressed compared to $\rm (SS)^{1j}$, but by how much requires explicit calculation.  

To take multi-jet recombination into account for the production of proton, we show more explicitly the terms in Eq.\ (\ref{34}), but still symbolically, 
\begin{eqnarray}
F_{qqq} = {\cal TTT + TTS + T(SS)}^{1j} + {\cal (SSS)}^{1j} + {\cal T(SS)}^{2j} +[{\cal  (SS)}^{1j}{\cal S]}^{2j} + {\cal (SSS)}^{3j}   \label{57}
\end{eqnarray}
Except for the first term that does not involve any $\rm S$, the other six terms are depicted by the six figures in Fig.\ 6, respectively. The first three figures have only 1-jet and are conventional. Figure 6 (d) corresponds to Eq.\ (\ref{55}) plus one thermal parton, so the equation for it is
\begin{eqnarray}
 \mathcal T \widehat{({\mathcal S\cal S})}^{2j}=\Gamma\int\frac{dp_1}{p_1}\frac{dp_2}{p_2}\frac{dp_3}{p_3}\mathcal T(p_1)\hat F_i(q_2)  S_i^q(p_2, q_2)\hat F_{i'}(q_3) S_{i'}^{q'}(p_3, q_3)R^p(p_1, p_2, p_3, p_T).  \label{58}
\end{eqnarray}
The last two figures can easily be obtained by straightforward generalization
\begin{eqnarray}
({\widehat{\mathcal {SSS}}})^{2j}&=&\Gamma\int\left[\prod\limits_{a=1}^3\frac{dp_a}{p_a}\right]\hat F_i(q_1) \Large\{S_i^q(p_1, q_1),  S_{i}^{q'}(p_2, q_1)\Large\}\nonumber \\
&&\times \hat F_{i'}(q_2) S_{i'}^{q''}(p_3, q_2)R^p(p_1, p_2, p_3, p_T),  \label{59} \\
({\widehat{\mathcal {SSS}}})^{3j}&=&\Gamma^2\int\left[\prod\limits_{a=1}^3\frac{dp_a}{p_a}\hat F_{i_a}(q_a)  S_{i_a}^{q_a}(p_a, q_a)\right]R^p(p_1, p_2, p_3, p_T).  \label{510}
\end{eqnarray}
Three-jet recombination is highly suppressed and will be neglected in the following.
 \begin{figure}[tbph]
\centering
\vspace*{-3cm}
\includegraphics[width=0.8\textwidth]{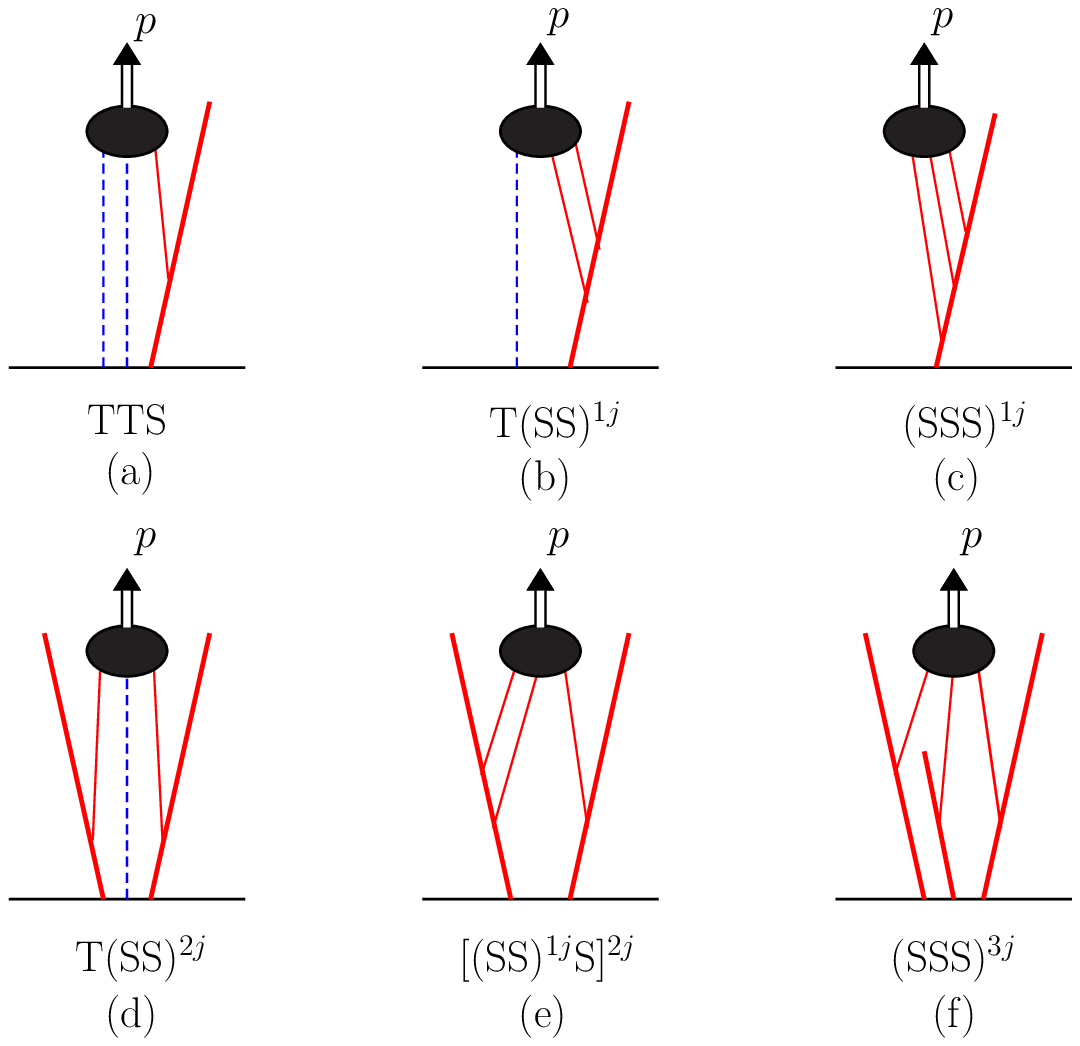}
\vspace*{-5cm}
\caption{(Color online) Diagrams showing the inclusive processes for proton production by recombination of partons with same line-types as in Fig.\ 5.}
\end{figure}

\section{Transverse Momentum Distributions of Hadrons}
We now calculate the $p_T$ distributions of $\pi, p, K$ and $\Lambda$ produced at $\eta\sim0$ and for 0-5\% centrality in Pb-Pb collisions at 2.76 TeV. They are based on the essential points discussed in the preceding sections, some of which have previously been applied to collisions at RHIC \cite{hy1, hz2}. Now we consider LHC without changing the basic formalism. Although we have studied the $p_T$ spectra at LHC before \cite{hz1}, it was, however, for a limited range of $p_T$ ($<5$ GeV/c) and was based on a simple assumption about momentum degradation, which we have subsequently found to be unrealistic as the $p_T$ range is extended to above 10 GeV/c. Our present treatment of momentum degradation, discussed in Sec.\ IV, enables us below to reproduce the data up to $p_T\sim20$ GeV/c, thus superseding the earlier parametrizations in \cite{hz1}. Nevertheless, we stress by reiterating that the basic equations are the same, as summarized in \cite{hz2}, except that a new $\gamma_g$ is to be adjusted to fit the data.

\subsection{Pion and proton production}
To be specific we consider the production of $\pi^+$
\begin{eqnarray}
{dN^{TT}_{\pi}\over p_Tdp_T} &=&\frac{C^2}{6}e^{-p_T/T} ,   \label{61}\\
{dN_{\pi}^{TS}\over p_Tdp_T} &=& {C\over p_T^3} \int_0^{p_T} dp_1 p_1e^{-p_1/T}
\left[{\cal S}^{u}(p_T-p_1) +{\cal S}^{\bar d}(p_T-p_1)\right] ,    \label{62} \\
{dN^{{SS}^{1j}}_{\pi}\over p_Tdp_T} &=& {1\over p_T} \int {dq\over q^2} \sum_i \hat{F}_i(q)D^{\pi}_i(p_T,q) ,   \label{63}\\
{dN_{\pi}^{{SS}^{2j}}\over p_Tdp_T} &=& {\Gamma\over p_T^3} \int_0^{\pt} dp_1  {\cal S}^{u}(p_1) {\cal S}^{\bar d}(p_T-p_1) .     \label{64} 
\end{eqnarray}
In the above the pion mass has been neglected. 

For proton whose mass is certainly not negligible, we replace $p^0$ in Eq.\ (\ref{32}) by the transverse mass $m_T^p=(m_p^2+p_T^2)^{1/2}$ for $\eta=0$. With the RF given in Eq.\ (\ref{38}), we have
\begin{eqnarray}
\frac{dN_p^{TTT}}{p_Tdp_T}=g_{st}^pg_pg_p'\frac{C^3p_T^2}{m_T^p}e^{-p_T/T}, \label{65}
\end{eqnarray}
where $g_p'=B(\alpha+2, \beta+2)B(\alpha+2, \alpha+\beta+4)$, $\alpha$ and $\beta$ being given after Eq.\ (\ref{38}), and
\begin{eqnarray}
{dN_p^{TTS}\over \pt d\pt}&=&{g_{st}^pg_p C^2\over m_T^p \pt^{2\alpha+\beta+3}} \int_0^{\pt} dp_1 \int_0^{\pt-p_1} dp_2\ e^{-(p_1+p_2)/T}  \nonumber \\
	&& \hspace{1cm} \times\left\{ (p_1p_2)^{\alpha+1}(\pt-p_1-p_2)^{\beta} {\cal S}^d(\pt-p_1-p_2)\right.\nonumber\\
&&\hspace{1cm}\left.+p_1^{\alpha+1}p_2^{\beta+1}(\pt-p_1-p_2)^{\alpha} {\cal S}^u(\pt-p_1-p_2)\right\}, \label{66}
\end{eqnarray}

\begin{eqnarray}
{dN_p^{{TSS}^{1j}}\over \pt d\pt}&=&{g_{st}^pg_p C\over m_T^p \pt^{2\alpha+\beta+3}} \int_0^{\pt} dp_1 \int_0^{\pt-p_1} dp_2\ e^{-p_1/T}  \nonumber \\
	&& \hspace{1cm} \times\left\{ p_1^{\beta+1}p_2^{\alpha}(\pt-p_1-p_2)^{\alpha} {\cal S}^{uu}(p_2,\pt-p_1-p_2)\right.\nonumber\\
&&\hspace{1cm}\left.+p_1(p_1p_2)^{\alpha}(\pt-p_1-p_2)^{\beta} {\cal S}^{ud}(p_2,\pt-p_1-p_2)\right\},\label{67}
\end{eqnarray}

\begin{eqnarray}
{dN_p^{{SSS}^{1j}}\over \pt d\pt}=\frac{1}{m_T^p}\int\frac{dq}{q^2}\sum\limits_i\hat F_i(q)D_i^p(p_T, q),\label{68}
\end{eqnarray}
where $\mathcal{S}^{qq}$ in Eq.\ (\ref{67}) is
\begin{eqnarray}
\mathcal{S}^{qq}(p_2, p_3)=\int\frac{dq}{q}\sum\limits_i\hat F_i(q){\rm S}_i^q(p_2, q){\rm S}_i^q(p_3, q-p_2).\label{69}
\end{eqnarray}
Equations (\ref{66})-(\ref{68}) correspond to Fig.\ 6(a)-(c). For 2-jet contributions in Fig.\ 6(d) and (e) we have
\begin{eqnarray}
{dN_p^{{TSS}^{2j}}\over \pt d\pt}&=&{g_{st}^pg_p  C\Gamma\over m_T^p \pt^{2\alpha+\beta+3}} \int_0^{\pt} dp_1 \int_0^{\pt-p_1} dp_2\ e^{-p_1/T}  \nonumber \\
	&& \hspace{1cm} \times\left\{ p_1^{\beta+1}p_2^{\alpha}(\pt-p_1-p_2)^{\alpha} {\cal S}^u(p_2) {\cal S}^{u}(\pt-p_1-p_2)\right.\nonumber\\
&&\hspace{1cm}\left.+p_1(p_1p_2)^{\alpha}(\pt-p_1-p_2)^{\beta} {\cal S}^u(p_2) {\cal S}^{d}(\pt-p_1-p_2)\right\}, \label{610}
\end{eqnarray}

\begin{eqnarray}
{dN_p^{{SSS}^{2j}}\over \pt d\pt}&=&{g_{st}^pg_p\Gamma \over m_T^p \pt^{2\alpha+\beta+3}} \int_0^{\pt} dp_1 \int_0^{\pt-p_1} dp_2  \nonumber \\
	&& \hspace{1cm} \times\left\{ p_1^{\beta}p_2^{\alpha}(\pt-p_1-p_2)^{\alpha} {\cal S}^d(p_1) {\cal S}^{uu}(p_2,\pt-p_1-p_2)\right.\nonumber\\
&&\hspace{1cm}\left.+(p_1p_2)^{\alpha}(\pt-p_1-p_2)^{\beta} {\cal S}^u(p_1) {\cal S}^{ud}(p_2,\pt-p_1-p_2)\right\}. \label{611}
\end{eqnarray}

The combination of all the equations (\ref{61})-(\ref{611}) collectively describe the production of pion and proton in the recombination model for hadronization at the final stage of the nuclear collision process where the medium density is low. 
As mentioned earlier in Sec.\ III, the thermal partons include the soft partons generated by hard and semihard partons as they traverse the medium and are assumed to have the thermal distribution described by  Eq.\ (\ref{35}) at the end of the deconfined phase. When those thermal partons are dilute enough and are ready for confinement through recombination, their local hadronization process is not  sensitive to  the collisional system in which the medium is created initially. 
The concept is consistent with the notion of universal hadrosynthesis where statistical study of hadron ratios has found universality independent of collision energy, analogous to water vapor condensing at 100$^{\circ}$C independent of how hot it has previously been.
Local process carries no information of the global properties, such as rapidity range and overall multiplicities, which depend on the collision energy. 
The \dis s we study are at mid-rapidity, so the increase of total multiplicity with energy that is due largely to the broadening of the rapidity plateau is not of concern here. Our interest is in the increase of $\left.dN/d\eta\right|_{\eta\sim 0}$ at LHC relative to RHIC, which we claim is related to the increase of $T$ and ${\cal S}^q(p_2)$ by  demonstrating that the observed spectra at LHC can be reproduced in the RM. 

Before we proceed to vary $C$ and $(\gamma_0, q_0)$ in Eq.\ (\ref{47}) to fit the data, it is necessary to remark that there are contributions from resonance decay that populate the  low \ppt\ region of the pion distribution. Those contributions are not accounted for above. We relinquish any attempt to calculate them, and add a term $u(\pt)$  to Eq.\ (\ref{61}), modifying it to
\bq
{dN^{TT}_{\pi}\over p_Tdp_T} &=&[1+u(\pt)]\frac{C^2}{6}e^{-p_T/T} ,  \label{612}
\eq
where $u(\pt)$ is to be determined by improving the fit of the region $\pt<2$ GeV/c while the overall \ppt\ distributions are fitted. The latter is achieved by using
\bq
C=23.2\hspace{0.05cm} \mbox{(GeV/c)$^{-1}$}, \hspace{0.5cm} \gamma_0=2.8, \hspace{0.5cm} q_0=7\ \mbox{GeV/c}. \label{613}
\end{eqnarray}

In Fig.\ 7 we show the results of our calculation of the pion distribution for $0<p_T<20$ GeV/c, exhibiting the different components  by different line types. The black line with black crosses is the sum of all four components without resonances and agrees with data from ALICE \cite{kslambda} very well for  $p_T>2.5$ GeV/c. The addition of the $u(\pt)$ term in Eq.\ (\ref{612}) to account for the resonances results in the solid black line that is in perfect agreement with the data for all \ppt; that term is
\bq
u(p_T)=3.95\times e^{-p_T/0.45}.  \label{614}
\eq
The total goes below the data points at $p_T>18$ GeV/c. Some further adjustment of $\gamma_g(q)$ at very high $q$ can repair that deficiency by raising $\rm SS^{1j}$ there, but that much fine tuning is not our interest  here, since our focus is on the interplay among the different components at low and intermediate $p_T$. 

Without changing any parameter we calculate the proton distribution that is shown in Fig.\ 8. It also agrees with the data \cite{Adam:2015kca} extremely well. Note that $\rm TTT$, $\rm TTS$, $\rm TSS^{1j}$ and $\rm SSS^{1j}$ components are all of similar magnitudes at $p_T\approx6$ GeV/c; together they lift the total to meet the data points. That is a feature that is unique among the hadronization models. It demonstrates the importance of minijets in the intermediate \ppt\ region.

 \begin{figure}[tbph]
\vspace*{-0.5cm}
\includegraphics[width=.8\textwidth]{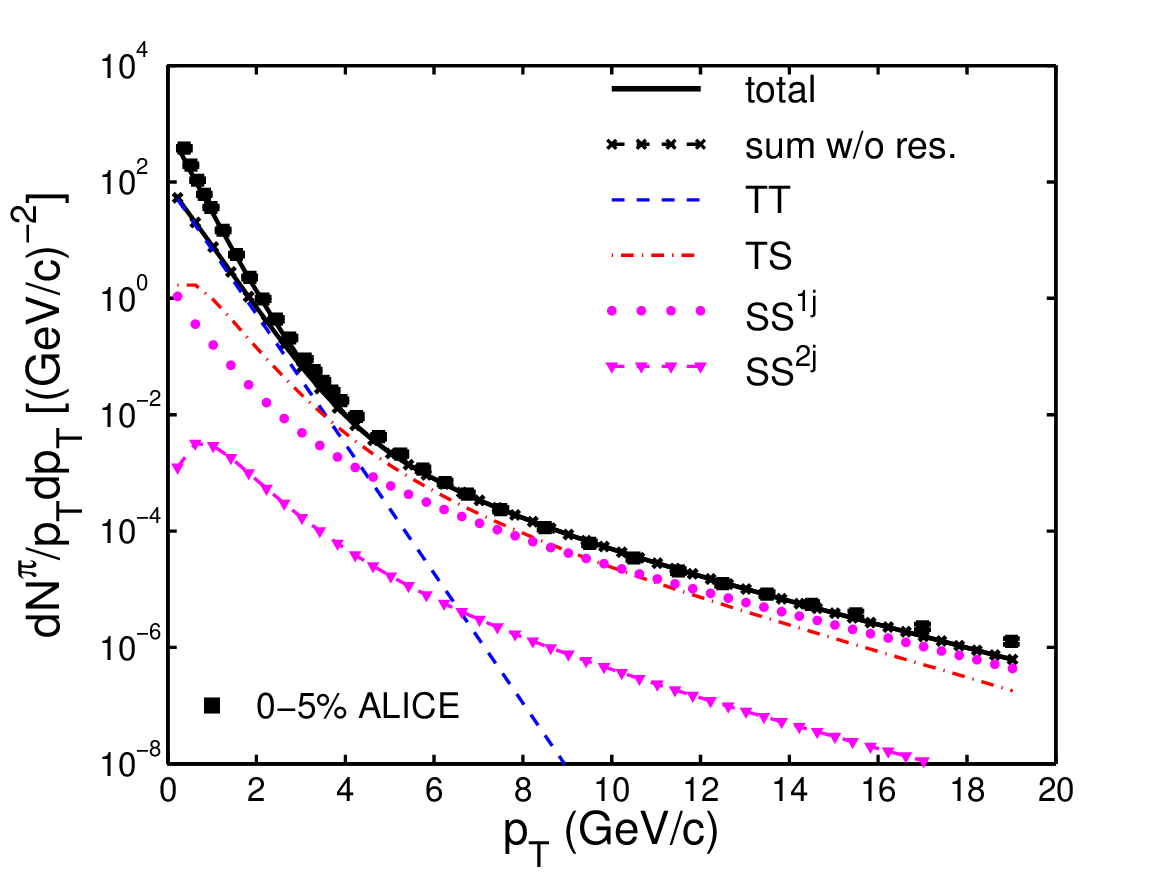}
\caption{(Color online) Transverse momentum distribution of pion produced in Pb-Pb collision at $\sqrt{s_{NN}}=2.76$ TeV. Data are from \cite{Adam:2015kca} for centrality 0-5\%. }
\end{figure}

 \begin{figure}[tbph]
\vspace*{-0.5cm}
\includegraphics[width=.8\textwidth]{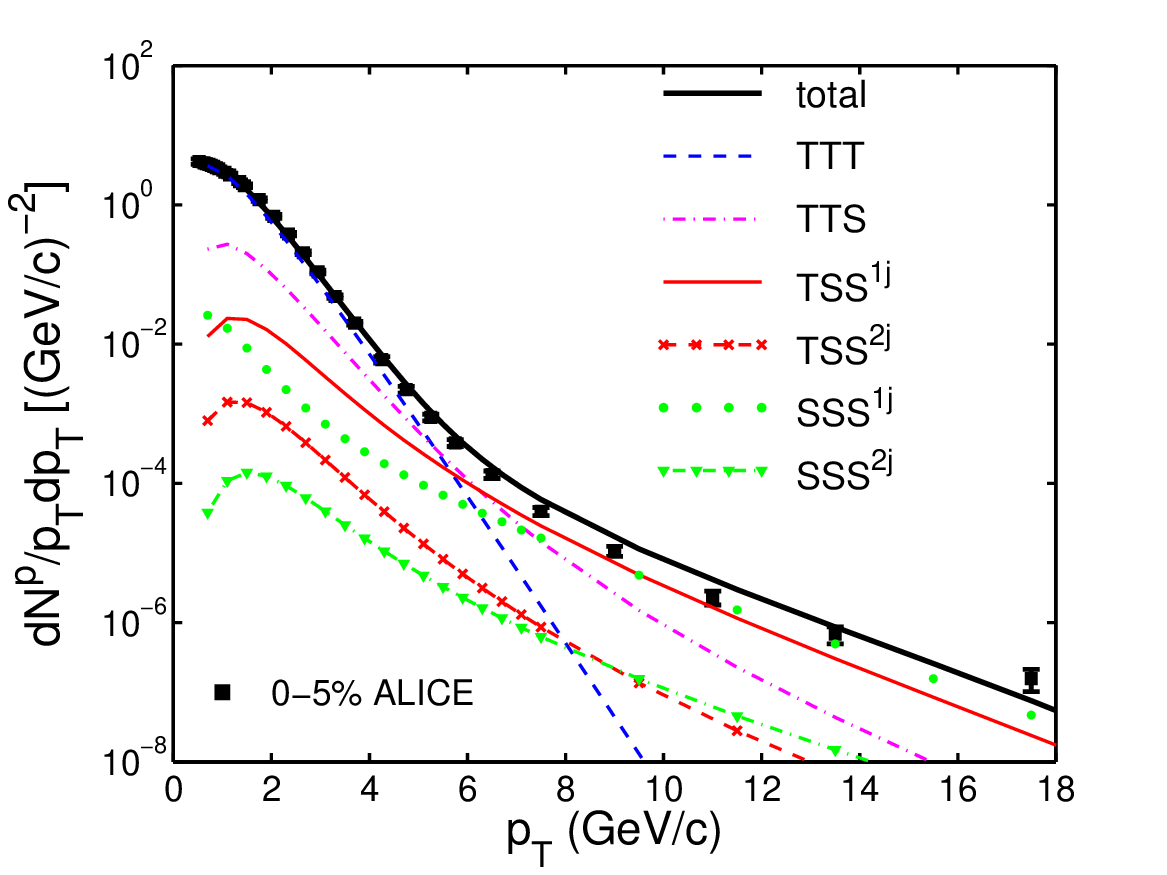}
\caption{(Color online) Transverse momentum distribution of proton produced in Pb-Pb collision at $\sqrt{s_{NN}}=2.76$ TeV. Data are from \cite{Adam:2015kca} for centrality 0-5\%. }
\end{figure}

With the results shown in Figs.\ 7 and 8 we regard our main objective as having been accomplished. It is non-trivial to reproduce the data in such a wide range of $p_T$ and it is remarkable that the main unadjustable input, i.e., the value  $T=0.39$ GeV/c given in Eq.\ (\ref{3.8}) and obtained in \cite{Hwa:2018qss} for baryons, works here for pions as well. 
 What we have obtained for $\gamma_0$ and $q_0$ in Eq.\ (\ref{613}) that are free parameters in the momentum degradation factor $\gamma_g(q)$ in Eq.\ (\ref{47}) 
 are good not only for $\pi$ and $p$ distributions, but also for all other particles, as we shall show below. That degradation factor is crucial in the description of the effects that the hard and semi-hard jets have on the spectra in the intermediate \ppt\ range. 
 Thus our result strongly supports the assertion that minijet  production plays a very important role in the structure of hadronic spectra. The corresponding shower partons have been exhibited already in Fig.\ 4 together with discussions on their dominance for $p_1>3$ GeV/c.

\subsection{$K$ and $\Lambda$ production}
Proceeding to the production of strange particles, we use the same formalism as for pion and proton, except that $s$ quark being more massive than the light quarks requires separate attention. For the thermal $s$ quarks we use the same distribution as in Eq.\ (\ref{35}) 
\begin{eqnarray}
\mathcal{T}^s(p_1)=C_sp_1e^{-p_1/T_s} \label{615}
\end{eqnarray}
but with different normalization factor $C_s$ and inverse slope $T_s$. The value of $T_s$  has been determined by the $\Omega$ spectrum in Ref. \cite{Hwa:2018qss} and is shown in Eq.\ (\ref{3.9}). Therefore, $C_s$ is the only parameter we adjust to fit the data. 
The $s$ shower parton distribution $\mathcal{S}^s(p_2)$ is as given in Eq.\ (\ref{36}) with the unintegrated SPD $S_i^j(z)$ determined from the FFs into $K$ and $\Lambda$ \cite{hy1,AKK08}. The degradation of $s$-quark momentum is taken to be the same as others, i.e., $\gamma_s=\gamma_q=\gamma_g/2$.

With the RF for kaon given in Ref.\ \cite{hy8,hy9} we have for the $K^+$ distributions
\begin{eqnarray}
{dN^{TT}_{K}\over p_Tdp_T} &=&
{12CC_s\over m_T^Kp_T^5} \int_0^{\pt} dp_1 p_1(\pt-p_1)^2p_1e^{-p_1/T}(p_T-p_1)e^{-(p_T-p_1)/T_s} ,     \label{616}\\
{dN_{K}^{TS}\over p_Tdp_T} &=& {12\over m_T^Kp_T^5} \int_0^{\pt} dp_1 p_1^2(\pt-p_1)^2  \nonumber  \\
&&\times \left[Ce^{-p_1/T}{\cal S}^{\bar s}(p_T-p_1,c) +C_s\left({\pt\over p_1}-1\right)e^{-(\pt-p_1)/T_s}{\cal S}^u(p_1)\right] ,    \label{617} \\
{dN^{{SS}^{1j}}_{K}\over p_Tdp_T} &=& {1\over m^K_T} \int {dq\over q^2} \sum_i \hat{F}_i(q)D^{K}_i(p_T,q)
,  \label{618}\\
{dN_{K}^{{SS}^{2j}}\over p_Tdp_T} &=& {12\Gamma\over m_T^Kp_T^5} \int_0^{\pt} dp_1 p_1(\pt-p_1)^2 {\cal S}^{u}(p_1) {\cal S}^{\bar s}(p_T-p_1) .    \label{619} 
\end{eqnarray}
With $C_s$ being the only adjustable parameter we obtain for 
\begin{eqnarray}
C_s=11\hspace{0.1cm}\mbox{(GeV/c)$^{-1}$} \label{620}
\end{eqnarray}
the distribution shown in Fig.\ 9. Evidently, the data from ALICE Collaboration \cite{kslambda} are well reproduced. Although $\mathcal{S}^s(p_1)$ is suppressed relative to $\mathcal{S}^u(p_1)$, the $\bar su$ recombination sustains the $\rm TS$ component. However, $\rm{SS}^{1j}$ is clearly much lower than that for pion in Fig.\ 7 at low $p_T$. 

For $\Lambda$ production we use Eq.\ (\ref{615}) again for the thermal $s$ quarks with the same  $C_s$ and $T_s$. Appendix A contains the explicit formulas for the distributions of the various components. Without any parameters to adjust, the experimental data \cite{kslambda} are reproduced very well in Fig.\ 10. The physics is clearly very much the same as for $\pi, p$ and $K$. 
We recall that the \ppt\ \dis\ of $\Lambda$ is well described by thermal partons only in Ref.\ \cite{Hwa:2018qss}, but only for $\pt < 5$ GeV/c. That is visible in Fig.\ 10. Here we have included also the contributions from shower partons, and our description of the spectrum has successfully  been extended to 11 GeV/c.
We stress that the momentum degradation parameters have not been changed so the hard parton and minijet distributions $\hat F_i(q)$ are the same as described in Sec.\ IV, independent of the hadrons produced. Thus the recombination model has enabled us to calculate the spectra of both the strange and non-strange hadrons at all $p_T$ in a universal formalism. 

 \begin{figure}[tbph]
\vspace*{-0.5cm}
\includegraphics[width=.8\textwidth]{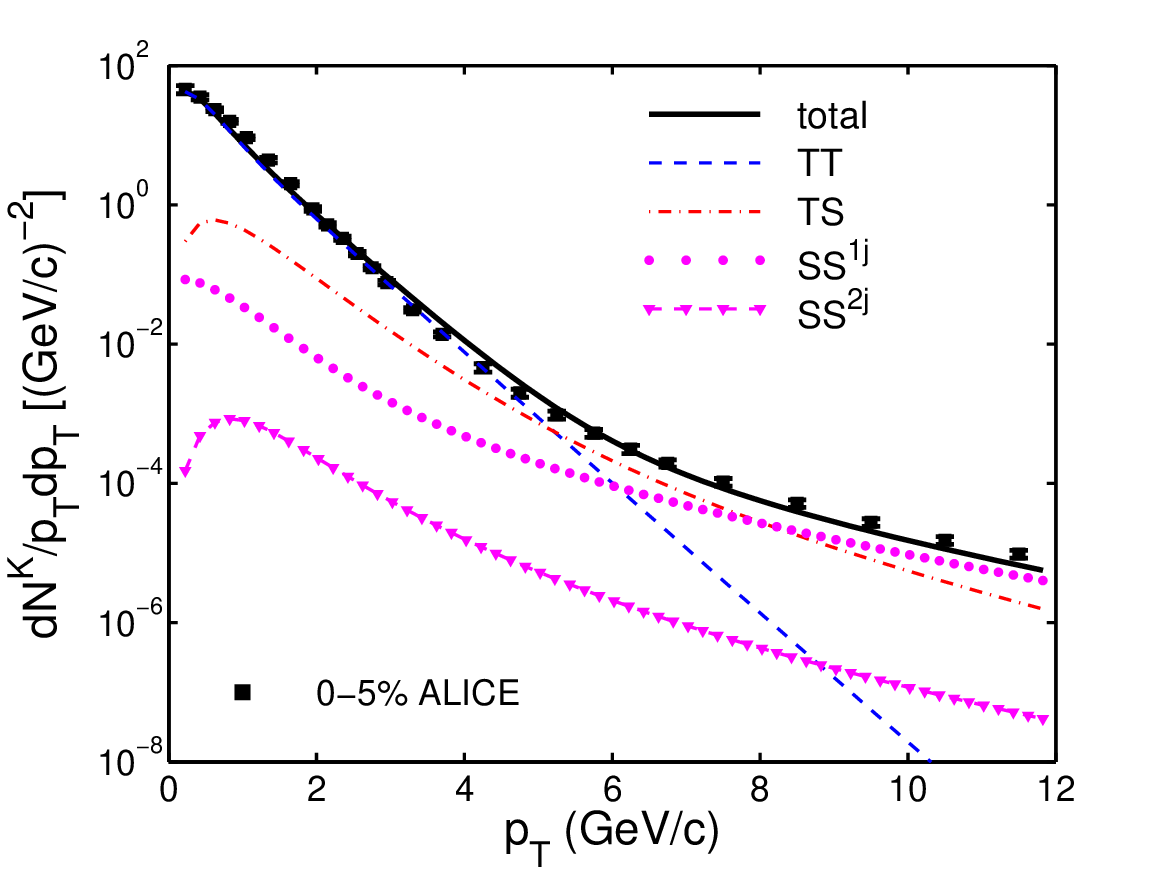}
\caption{(Color online) Transverse momentum distribution of kaon produced in Pb-Pb collision at $\sqrt{s_{NN}}=2.76$ TeV. Data are from \cite{kslambda} for centrality 0-5\%. }
\end{figure}

 \begin{figure}[tbph]
\vspace*{-0.5cm}
\includegraphics[width=.8\textwidth]{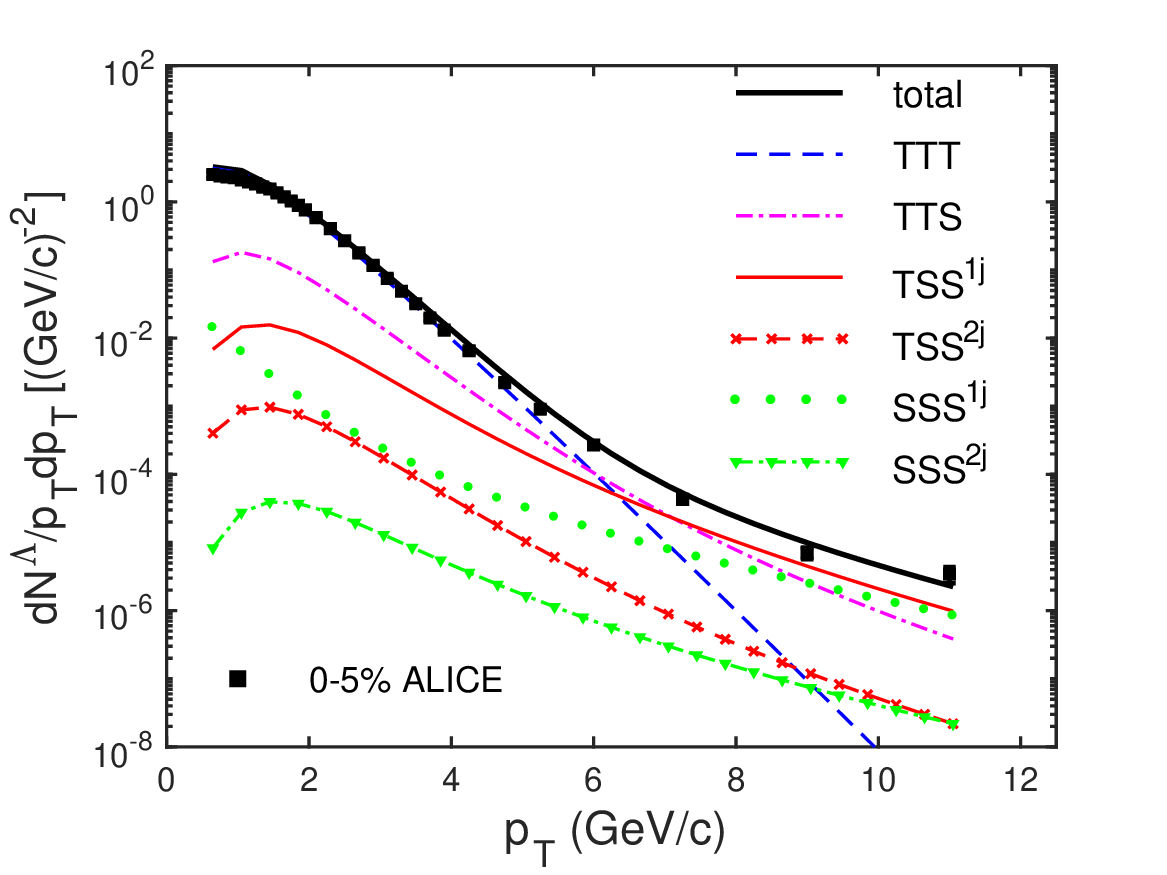}
\caption{(Color online) Transverse momentum distribution of $\Lambda$ produced in Pb-Pb collision at $\sqrt{s_{NN}}=2.76$ TeV. Data are from \cite{kslambda} for centrality 0-5\%. }
\end{figure}

\section{Multi-strange hyperons and meson}
We complete our investigation of hadron production by considering $\Xi$, $\Omega$ and $\phi$ . Apart from different quark contents of those particles, the physics of hadronization through recombination is the same as before. Since they cannot be used either as target on beam particles, their wave functions in terms of momentum fractions of constituent quarks are not known as firmly as we do with $\pi, K$ and $p$. Furthermore, there is the question of the probability for more than one strange quark to find one another to recombine. As the system expands, the plasma gets out of chemical equilibrium first because $gg\to s\bar s$ and $q\bar q\to s\bar s $ processes become less frequent than their reverses on account of $m_s>m_q>m_g$. Thus the density of $s$ quarks becomes lower. 
The language used above is that of the conventional interpretation of the expanding medium getting out of chemical equilibrium. We need not subscribe to the details of that description, while still adhering to the qualitative physical picture of the system that has general validity. Thus we proceed in the same manner as we have for $\pi$ and $p$.
For a single $s$ quark to hadronize at late time  there are abundant light quarks in the neighborhood of the $s$ quark to form $K$ and $\Lambda$. However, for multi-strange hadron to form, the probability of $ss$, $sss$ or $s\bar s$ to be in close proximity of one another at late time is reduced, when the density of $s$ quark is lower than that of light quarks. 
If at earlier time $\Xi$, $\Omega$ and $\phi$ are formed at higher density, their survival in the medium is suppressed due to their dissociation through interaction with the plasma that is still active. Thus in either case the rate of multi-strange hadron production is lower. We cannot predict that rate in the recombination model, so an adjustable parameter will be used to fit the overall normalization. On the other hand, the density of shower partons arising from hard and semihard partons is independent of the final hadrons formed, so we can still use our formalism to calculate the various components of the $p_T$ distributions. 

The detail equations for $\Xi$ and $\Omega$ formations are given in Appendices B and C, respectively. The only free parameter we use in each case is $g_h$. For best fit we obtain
\begin{eqnarray}
\hspace{0.5cm}g_{\Xi}=0.03,\hspace{0.3cm}g_{\Omega}=0.01, \label{71}
\end{eqnarray}
The results are shown in Figs.\ 11 and 12, reproducing the data very well. There are, however, some differences in the strengths of different components, even though the shower partons are the same in all cases.

 \begin{figure}[tbph]
\includegraphics[width=.8\textwidth]{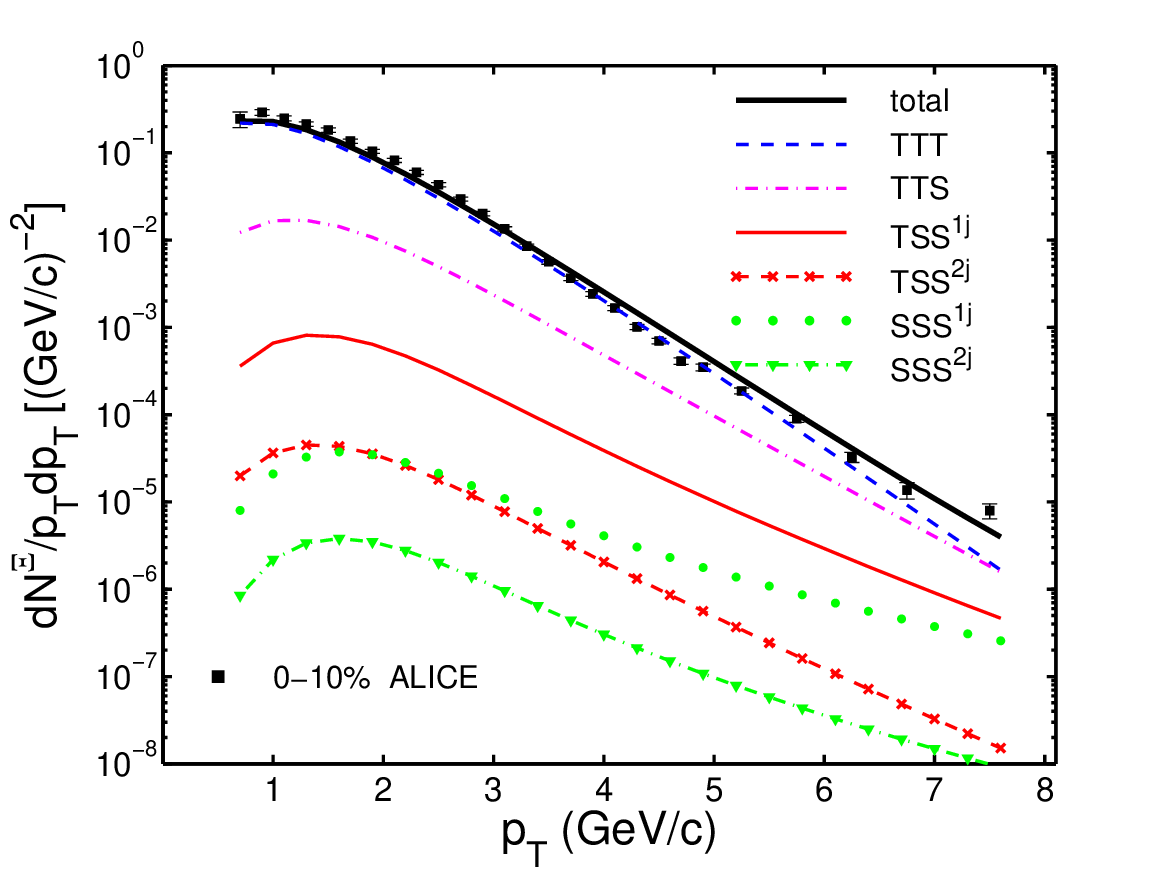}
\caption{(Color online) Transverse momentum distribution of $\Xi$ produced in Pb-Pb collision at $\sqrt{s_{NN}}=2.76$ TeV. Data are from \cite{ba2} for centrality 0-10\%. }
\end{figure}

 \begin{figure}[tbph]
\vspace*{-0.5cm}
\includegraphics[width=.8\textwidth]{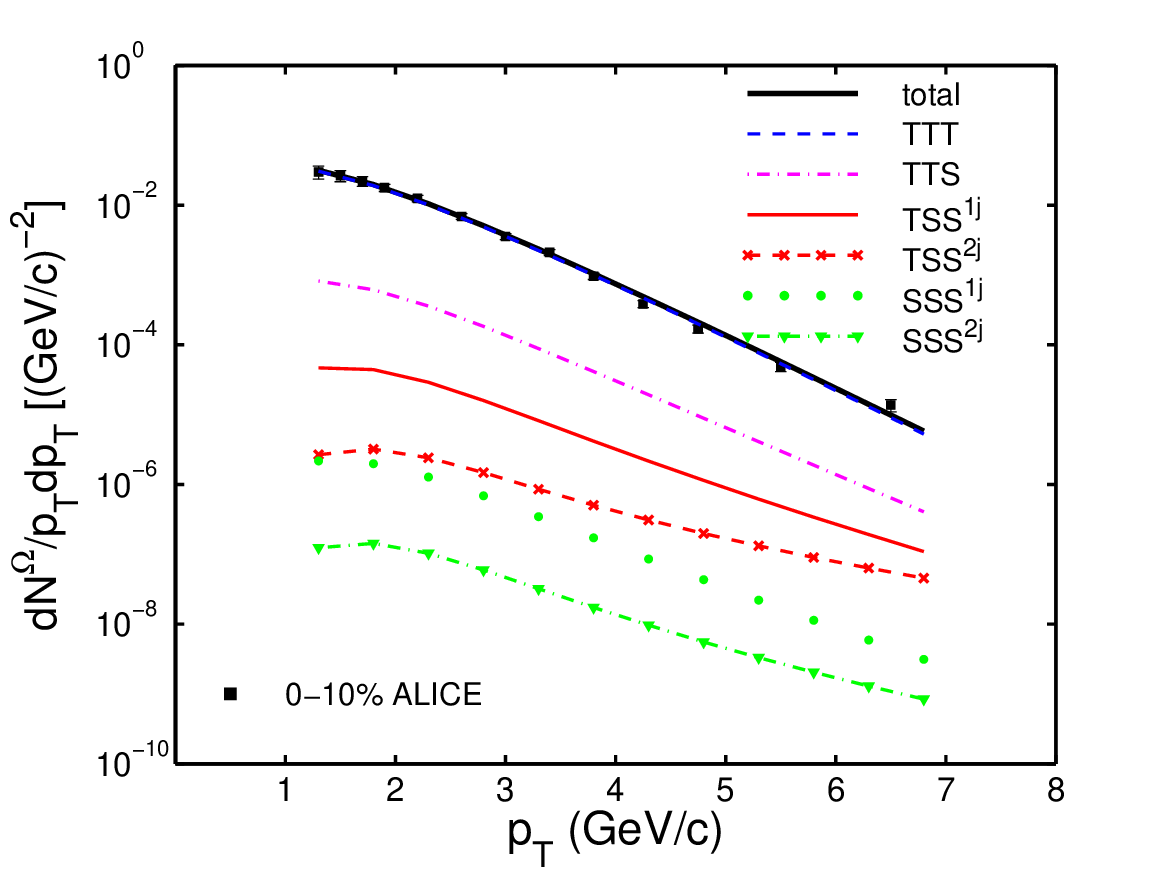}
\caption{(Color online) Transverse momentum distribution of $\Omega$ produced in Pb-Pb collision at $\sqrt{s_{NN}}=2.76$ TeV. Data are from \cite{ba2} for centrality 0-10\%. }
\end{figure}

What is most noticeable about the $\Xi$ distributions is that the $\rm{TTT}$ component dominates the whole spectrum for $p_T<7$ GeV/c and that $\rm{TSS}$ and $\rm{SSS}$ components are much lower. The relative strengths of those components are  unlike the situation with proton and $\Lambda$. Whereas the $\rm S$ in $\rm{TTS}$ can be non-strange, $\rm{TSS}$ must have at least one $s$ in  the $\rm SS$, and $\rm{SSS}$ must have two $s$ quarks. Since $S^s$ is suppressed compared to $S^q$, the ordering of $\rm{TTS}$, $\rm{TSS}$ and $\rm{SSS}$ is evident in Fig.\ 11. It is because the largest shower component $\rm{TTS}$ does not become important until \ppt\ exceeds 7 GeV/c that we could find in Ref.\ \cite{Hwa:2018qss} the simple exponential behavior in the data.

For $\Omega$ production shown in Fig.\ 12, similar remarks about the ordering of the various components can be made as for $\Xi$. One notable difference is that this time even $\rm{TTS}$ is suppressed relative to $\rm{TTT}$ at all $p_T$. That is because every coalescing quark for $\Omega$ must be strange, so $S^s$ in $\rm{TTS}$ lowers its magnitude relative to $\rm{TTT}$. Herein lies a very interesting point that was noticed several years ago even in RHIC data \cite{ja,hw1}. The $p_T$ distribution of $\Omega$ is exponential 
(apart from the prefactor $p_T^2/m_T^\Omega$ in Eq.\ (\ref{D1})) 
without any power-law up-bending at high $p_T$. It means that $\Omega$ is produced thermally even at $p_T\sim 6$ GeV/c without any contribution from parton fragmentation, which is the usual mechanism considered in pQCD.

Lastly, we consider the production of $\phi$, for which the equations are given in Appendix D. No light quarks are involved in the formation of both $\Omega$ and $\phi$. We still use the same value of $T_s=0.51$ GeV for $\phi$. By varying $g_{\phi}$ only for the overall normalization, we obtain the result shown in Fig.\ 13 for $g_\phi=0.432$. The underlying components are very similar to those for $\Omega$, namely: TT dominates over TS, while SS (whether 1j or 2j) is nearly 3 orders of magnitudes farther down.

 \begin{figure}[tbph]
\vspace*{-0.5cm}
\includegraphics[width=.8\textwidth]{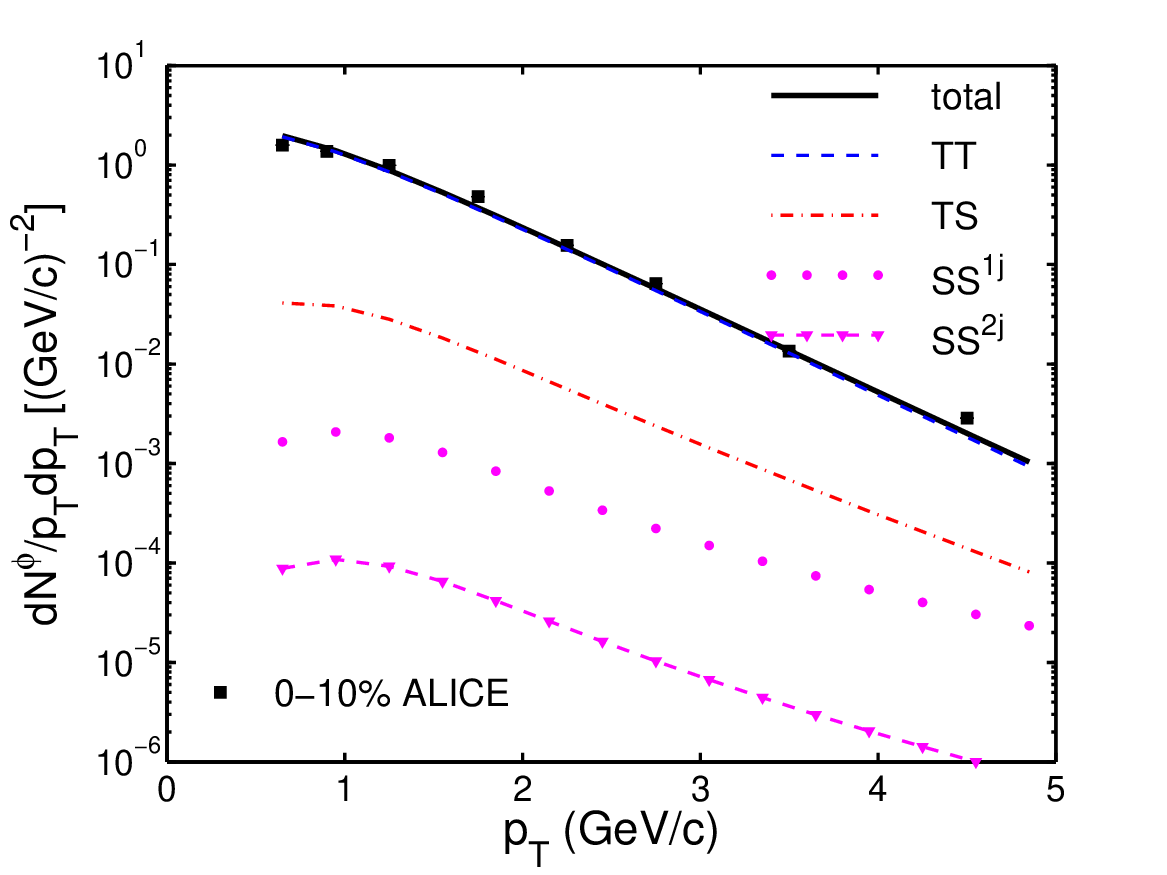}
\caption{(Color online) Transverse momentum distribution of $\phi$ produced in Pb-Pb collision at $\sqrt{s_{NN}}=2.76$ TeV. Data are from \cite{ba2} for centrality 0-10\%. }
\end{figure}

The small value of $g_\phi$ relative to $g_\pi$ being 1 in Eq.\ (\ref{37}) is an indication of  quarkonium suppression after $\phi$ is formed at a time much earlier than $\pi$, when the density of $s$ (and $\bar s$) is higher. As is the case with $J/\psi$ suppression, $\phi$ experiences the effects of dissociation by the plasma as it traverses the remaining portion of the medium before it completely hadronizes. The value of $g_\phi$ depends on aspects of the process that are not included in the formalism discussed in this paper, and therefore cannot be predicted. The same remarks can be made for the formation of $\Xi$ and $\Omega$, for which $g_\Xi$ and $g_\Omega$ are quite small in Eqs.\ (\ref{71}).

\section{Conclusion}
We have made a thorough investigation of the production of all identified hadrons in Pb-Pb collisions at LHC in a universal formalism that is applicable to the entire \ppt\ spectra of all particles observed in central collisions. We
have studied all  components of thermal- and shower-parton recombination that contribute to the hadronization processes in all sectors of strangeness. In the wide ranges of \ppt\ in which the strange baryons are observed, their \ppt\ \dis s are found to be exponential with large values of inverse slopes.  That led us to realize the importance of the effects of  semihard and hard partons on the density of soft partons in the medium before hadronization. 
That is, parton energy loss in the medium must increase the thermal energies of the soft partons, light or strange, and affect the nature of the whole spectra of all particles produced.

The degradation of momenta of hard and semihard partons is treated in a way that uses two free parameters, which are determined by fitting the high-$p_T$ distribution of the pion. The resultant shower-parton distributions of $q$ and $s$ quarks are then used to supplement the thermal partons 
in calculating the spectra of all identified hadrons ($\pi, K, p, \Lambda$, $\Xi$, $\Omega$ and $\phi$). They agree well with the data for all $p_T$ up to 20 GeV/c. The description not only establishes a consistent scheme for treating the hadronization process of a quark-gluon plasma at LHC, but also points out the importance of minijets, which affect the \ppt\ spectra in two ways. One is, of course, the explicit contributions of shower partons in recombination. The other is the indirect effects of hard and semihard partons on the thermal partons that include the soft partons generated by them in the medium, resulting in increased values of the inverse slopes $T_q$ and $T_s$ in the distributions $\mathcal{T}^{q,s}(p_1)$ determined in Ref. \cite{Hwa:2018qss}. 

 The thermal partons dominate the low-$p_T$ region of all particles produced. 
For pion and protons the various combinations of the light quarks in the thermal and shower partons make important contributions in different regions of \ppt.
For $\phi$ and $\Omega$, the $s$ quarks in the shower are suppressed, so $\rm{TS}$ and $\rm{TTS}$ are lower than $\rm{TT}$ and $\rm{TTT}$, respectively. The other particles ($K,\Lambda$ and $\Xi$) with less strangeness contents are in the intermediate situation.
The contributions from the recombination of shower partons belonging to separate but nearby jets turn out generally to be negligible.
The formalism interconnects the hadronization of all soft and semihard partons,  and  is therefore highly constrained by the observed hadronic data.  The fact that we are able to explain  the production of $\phi$ and $\Omega$ by means of the recombination of thermal partons only for $p_T$  up to 6.5 GeV/c raises question on alternative approaches that do not rely on recombination, since $p_T> 3$ GeV/c is too high for hydrodynamics and $\phi$ and $\Omega$ are too abundantly produced for fragmentation.

Concerning azimuthal  anisotropy in non-central collisions, the usual explanation is that the azimuthal harmonics are due to the flow effects of the fluctuations of the initial configuration of the collision system. If, however, the non-flow effects such as minijets are important, the fluid treatment would be inadequate on the one hand, and our approach is in need of suitable treatment to be convincing on the other. For Au-Au collisions at 200 GeV, we have shown that the azimuthal harmonics can be obtained by taking into account the azimuthal dependence of minijet and the related ridge effect \cite{hz3}. Now for Pb-Pb collisions at 2.76 TeV we have only investigated the case of central collisions here. To extend the study to non-central collisions is, of course, the natural problem to pursue next. How minijets influence the azimuthal asymmetry will undoubtedly be a major area of investigation. The consideration described here represents only the first, but significant, step toward understanding the physics of hadronization at LHC.

\begin{appendix}
 
 \section{$p_T$ Distribution of $\Lambda$ at LHC}
 The $p_T$ distribution of $\Lambda$ is very similar to that of proton except for the replacement of a $u$ quark by an $s$ quark. The thermal and shower parton distributions for $s$ are different from those for $u$, and the RF for $\Lambda$ is different from that for $p$. For $\mathcal{T}^s(p_1)$ we use the same form as Eq. (\ref{615}), with $T_s$ being given by (\ref{3.9}). $\mathcal{S}^s(p_2)$ is the same as used for $K$ production in Sec. VI-B. The RF for $\Lambda$ has the same form as Eq. (\ref{38}) for proton but with $\alpha=1$ and $\beta=2$ in a problem on strange particle production at RHIC considered in Ref. \cite{hy9}. We simply list the equation below for the various components.
 
 \begin{eqnarray}
{dN_{\Lambda}^{TTT}\over \pt d\pt}&=&{g_{st}^{\Lambda}g_{\Lambda}C^2C_s\over m_T^{\Lambda} \pt^{2\alpha+\beta+3}} \int_0^{\pt} dp_1 \int_0^{\pt-p_1} dp_2 \nonumber \\
	&& \hspace{1cm} \times (p_1p_2)^{\alpha+1}\ e^{-(p_1+p_2)/T}(p_T-p_1-p_2)^{\beta+1}\ e^{-(p_T-p_1-p_2)/T_s}, \label{B1}
\end{eqnarray}
	
\begin{eqnarray}
{dN_{\Lambda}^{TTS}\over \pt d\pt}&=&{g_{st}^{\Lambda}g_{\Lambda} \over m_T^{\Lambda} \pt^{2\alpha+\beta+3}} \int_0^{\pt} dp_1 \int_0^{\pt-p_1} dp_2 \nonumber \\
	&& \hspace{1cm} \times\left\{C^2p_1p_2\ e^{-(p_1+p_2)/T}  (p_1p_2)^{\alpha}(\pt-p_1-p_2)^{\beta}{\cal S}^s(\pt-p_1-p_2)\right.\nonumber\\
&&\hspace{1cm}\left.+CC_sp_1\ e^{-p_1/T}p_2\ e^{p_2/T_s} p_1^{\alpha}p_2^{\beta}(\pt-p_1-p_2)^{\alpha} {\cal S}^u(\pt-p_1-p_2)\right\},
\label{B2}
\end{eqnarray}

\begin{eqnarray}
{dN_{\Lambda}^{TSS^{1j}}\over \pt d\pt}&=&{g_{st}^{\Lambda}g_{\Lambda} \over m_T^{\Lambda} \pt^{2\alpha+\beta+3}} \int_0^{\pt} dp_1 \int_0^{\pt-p_1} dp_2 \nonumber \\
	&& \hspace{1cm} \times\left\{  C_sp_1\ e^{-p_1/T_s} p_1^{\beta}p_2^{\alpha}(\pt-p_1-p_2)^{\alpha} {\cal S}^{ud}(p_2,\pt-p_1-p_2)\right.\nonumber\\
&&\hspace{1cm}\left.+Cp_1\ e^{-p_1/T} (p_1p_2)^{\alpha}(\pt-p_1-p_2)^{\beta} {\cal S}^{ds}(p_2,\pt-p_1-p_2)\right\},
\label{B3}
\end{eqnarray}

\begin{eqnarray}
{dN_{\Lambda}^{{TSS^{2j}}}\over \pt d\pt}&=&{g_{st}^{\Lambda}g_{\Lambda} \Gamma\over m_T^{\Lambda} \pt^{2\alpha+\beta+3}} \int_0^{\pt} dp_1 \int_0^{\pt-p_1} dp_2\   \nonumber \\
	&& \hspace{1cm} \times\left\{ C_sp_1e^{-p_1/T_s}p_1^{\beta}p_2^{\alpha}(\pt-p_1-p_2)^{\alpha}{\cal S}^{u}(p_2){\cal S}^{d}(p_T-p_1-p_2) \right.\nonumber\\
&&\hspace{1cm}\left.+Cp_1e^{-p_1/T}(p_1p_2)^{\alpha}(\pt-p_1-p_2)^{\beta}{\cal S}^{d}(p_2){\cal S}^{s}(p_T-p_1-p_2) \right\},
\label{B4}
\end{eqnarray}

\begin{eqnarray}
{dN_{\Lambda}^{{SSS}^{1j}}\over \pt d\pt}=\frac{1}{m_{\Lambda}^T}\int\frac{dq}{q^2}\sum\limits_i\hat F_i(q)D_i^{\Lambda}(p_T, q),
\label{B5}
\end{eqnarray}

\begin{eqnarray}
{dN_{\Lambda}^{{SSS^{2j}}}\over \pt d\pt}&=&{g_{st}^{\Lambda}g_{\Lambda}\Gamma\over m_T^{\Lambda} \pt^{2\alpha+\beta+3}} \int_0^{\pt} dp_1 \int_0^{\pt-p_1} dp_2 \nonumber \\
	&& \hspace{1cm} \times\left\{ p_1^{\beta}p_2^{\alpha}(\pt-p_1-p_2)^{\alpha}{\cal S}^{s}(p_1) {\cal S}^{ud}(p_2,\pt-p_1-p_2)\right.\nonumber\\
&&\hspace{1cm}\left.+(p_1p_2)^{\alpha}(\pt-p_1-p_2)^{\beta} {\cal S}^{u}(p_1){\cal S}^{ds}(p_2,\pt-p_1-p_2)\right\},
\label{B6}
\end{eqnarray}
The statistical factor is $g_{st}^{\Lambda}=1/4$, and the prefactor from RF is $g_{\Lambda}=[B(\alpha+1, \alpha+\beta+2)B(\alpha+1, \beta+1)]^{-1}$. The corresponding FFs, $D_i^{\Lambda}(z)$, are given by AKK \cite{AKK08} by fitting the data at next-leading-order (NLO).

  \section{$p_T$ Distribution of $\Xi$ at LHC}
  
  For the recombination of $dss$ to form $\Xi$ we make the simplifying assumption that the RF is proportional to $\delta$-functions, 
  \begin{eqnarray}
  R^{\Xi}_{dss}(p_1,p_2,p_3,p_T) = g_{\Xi}p_1p_2p_3\prod\limits_{i=1}^3\delta(p_i-p_T/3) .
  \end{eqnarray}
$g_{\Xi}$ is an unknown numerical factor that summarizes the probability of the three quarks coalescing with each having 1/3 of the $\Xi$ momentum. It is the only parameter to be determined by fitting the normalization of the data.  With the above RF it is straightforward to write the distributions
  
 \begin{eqnarray}
{dN^{TTT}_{\Xi}\over p_Tdp_T} = {g_{\Xi}CC_s^2p_T^2\over 27m^{\Xi}_T}e^{-p_T/3T}e^{-2p_T/3T_s} ,  \label{C1}
\end{eqnarray}

\begin{eqnarray}
{dN^{TTS}_{\Xi}\over p_Tdp_T} = {g_{\Xi}p_T\over 9m^{\Xi}_T}\left\{{CC_sSe^{-\frac{p_T} {3} (\frac{1}{T}+\frac{1}{T_s})}} \mathcal S^s(p_T/3)+C_s^2e^{-2p_T \over 3T_s}\mathcal S^u(p_T/3)\right\} ,  \label{C2}
\end{eqnarray}

\begin{eqnarray}
{dN^{{TSS}^{1j}}_{\Xi}\over p_Tdp_T} =  {g_{\Xi}\over 3m^{\Xi}_T}\left\{Ce^{-\frac{p_T} {3T} } \mathcal S^{ss}(p_T/3, p_T/3)+C_se^{-\frac{p_T} {3T_s} }\mathcal S^{us}(p_T/3, p_T/3)\right\} ,  \label{C3}
\end{eqnarray}

\begin{eqnarray}
{dN^{{TSS}^{2j}}_{\Xi}\over p_Tdp_T} = {g_{\Xi}\Gamma\over 3m^{\Xi}_T}\left\{Ce^{-\frac{p_T} {3T} } \mathcal S^{s}(p_T/3,)\mathcal S^{s}(p_T/3)+C_se^{-\frac{p_T} {3T_s} }\mathcal S^{u}(p_T/3)\mathcal S^{s}(p_T/3)\right\} ,  \label{C4}
\end{eqnarray}

\begin{eqnarray}
{dN^{{SSS}^{1j}}_{\Xi}\over p_Tdp_T} = {g_{\Xi}\over {p_Tm^{\Xi}_T}} \mathcal S^{uss}(p_T/3, p_T/3, p_T/3) ,  \label{C5}
\end{eqnarray}

\begin{eqnarray}
{dN^{{SSS}^{2j}}_{\Xi}\over p_Tdp_T} ={ g_{\Xi}\Gamma\over p_Tm^{\Xi}_T}\Big\{\mathcal S^{u}(p_T/3) S^{ss}(p_T/3, p_T/3) +\mathcal S^{s}(p_T/3) S^{us}(p_T/3, p_T/3)\Big\} ,  \label{C6}
\end{eqnarray}

where
\begin{eqnarray}
\mathcal{S}^{dss}(p_1,p_2,p_3)=\int\frac{dq}{q}\sum\limits_i\hat F_i(q)S_i^u(p_1,q)S_i^s(p_2,q-p_1)S_i^s(p_3,q-p_1-p_2).   \label{C7}
\end{eqnarray} 
  
   \section{$p_T$ Distribution of $\Omega$ at LHC}
The RF for $\Omega$ can be found in Ref. \cite{hy9}, as it is for $\Xi$, 
\begin{eqnarray}
 R^{\Omega}_{sss}(p_1,p_2,p_3,p_T) = g_{\Omega}p_1p_2p_3\prod\limits_{i=1}^3\delta(p_i-p_T/3)
  \end{eqnarray}
$g_{\Omega}$ is a factor similar to $g_{\Xi}$ described in Appendix B. The distributions for the different components are simplest of all baryons, since all constituent quarks are the same. We have
  \begin{eqnarray}
{dN^{TTT}_{\Omega}\over p_Tdp_T} = {g_{\Omega}C_s^3p_T^2\over 27m^{\Omega}_T}e^{-p_T/T_s} ,  \label{D1}
\end{eqnarray}

\begin{eqnarray}
{dN^{TTS}_{\Omega}\over p_Tdp_T} = {g_{\Omega}C_s^2p_T\over 9m^{\Omega}_T}e^{-2p_T/3T_s} \mathcal S^s(p_T/3) ,    \label{D2}
\end{eqnarray}

\begin{eqnarray}
{dN^{{TSS}^{1j}}_{\Omega}\over p_Tdp_T} = {g_{\Omega}C_s\over 3m^{\Omega}_T}e^{-p_T/3T_s}\mathcal S^{ss}(p_T/3,p_T/3)  ,  \label{D3}
\end{eqnarray}

\begin{eqnarray}
{dN^{{TSS}^{2j}}_{\Omega}\over p_Tdp_T} ={ g_{\Omega}C_s\Gamma\over 3m^{\Omega}_T}e^{-p_T/3T_s}  \mathcal S^s(p_T/3) \mathcal S^s(p_T/3) ,  \label{D4}
\end{eqnarray}

\begin{eqnarray}
{dN^{{SSS}^{1j}}_{\Omega}\over p_Tdp_T} = {g_{\Omega}\over {p_Tm^{\Omega}_T}} \mathcal S^{sss}(p_T/3, p_T/3, p_T/3) ,  \label{D5}
\end{eqnarray}

\begin{eqnarray}
{dN^{{SSS}^{2j}}_{\Omega}\over p_Tdp_T} = {g_{\Omega}\Gamma\over p_Tm^{\Omega}_T} \mathcal S^s(p_T/3)\mathcal S^{ss}(p_T/3,p_T/3). \label{D6}
\end{eqnarray} 
Apart from the prefactor that involves $p_T^2/m_T^{\Omega}$, the $\rm{TTT}$ term is a pure exponential. If it is dominant, then the $p_T$ dependence of Eq.\ (\ref{D1}) is a direct test of the validity of our description of $\Omega$ production.
   
    \section{$p_T$ Distribution of $\phi$ at LHC}
   As it is for $\Omega$, the distributions for $\phi$ is simple when the RF is taken to be \cite{hy9}
\begin{eqnarray}
  R^{\phi}_{s\bar{s}}(p_1,p_2,p_T) = g_{\phi}p_1p_2\prod\limits_{i=1}^2\delta(p_i-p_T/2)
  \end{eqnarray}
$g_{\phi}$ is an unknown factor as with $g_{\Xi}$ and $g_{\Omega}$. One gets
   
 \begin{eqnarray}
{dN^{TT}_{\phi}\over p_Tdp_T} = {g_{\phi}C_s^2p_T\over 4m^{\phi}_T}e^{-p_T/T_s} ,    \label{E1}
\end{eqnarray}

\begin{eqnarray}
{dN^{TS}_{\phi}\over p_Tdp_T} = {g_{\phi}C_s\over 2m^{\phi}_T}e^{-p_T/2T_s} \mathcal S^s(p_T/2) , \label{E2}
\end{eqnarray}

\begin{eqnarray}
{dN^{{SS}^{1j}}_{\phi}\over p_Tdp_T} = {g_{\phi} \over p_Tm^{\phi}_T}\mathcal S^{s\bar s}(p_T/2, p_T/2) ,  \label{E3}
\end{eqnarray}

\begin{eqnarray}
{dN^{{SS}^{2j}}_{\phi}\over p_Tdp_T} = {g_{\phi} \Gamma\over p_Tm^{\phi}_T}\mathcal S^s(p_T/2)\mathcal S^{\bar s}(p_T/2) .  \label{E4}    
\end{eqnarray}      
  \end{appendix}

 \section*{Acknowledgment}
This work was supported,  in part,  by the NSFC of China under Grant No.\ 11205106.

\end{document}